\documentclass[modern]{aastex63}

\usepackage{chemformula}

\usepackage{CJK}
\newcommand\mps{m~s$^{-1}$}

\newcommand\mpscc{molecules~s$^{-1}$~cm$^{-2}$}

\newcommand\afrho[1][\theta]{\mbox{$A(#1)f\rho$}}
\newcommand\afr{\mbox{$Af\rho$}}

\newcommand\rh{\mbox{$r_{\mathrm{h}}$}}
\newcommand\inv[2][1]{$\textrm{#2}^{-#1}$}
\newcommand\timesten[1]{\mbox{$\times10^{#1}$}}
\newcommand\specgrad[2]{$S_{#1}=#2$\% per 100 nm}

\newcommand\hst{\textit{HST}}
\newcommand\hubble{\textit{Hubble Space Telescope}}
\newcommand\rosetta{\textit{Rosetta}}
\newcommand\di{\textit{Deep Impact}}
\newcommand\stardust{\textit{Stardust}}

\received{September 30, 2020}
\revised{April 19, 2021}
\accepted{May 03, 2021}
\submitjournal{The Planetary Science Journal}

\shortauthors{Kelley et al.}
\graphicspath{{./}{figures/}}

\begin{document}
\begin{CJK*}{UTF8}{gbsn}

  \title{Six Outbursts of Comet 46P/Wirtanen}

  \author[0000-0002-6702-7676]{Michael S. P. Kelley}
  \affil{Department of Astronomy, University of Maryland, College Park, MD 20742-0001, USA}
  \email{msk@astro.umd.edu}

  \author{Tony L. Farnham}
  \affil{Department of Astronomy, University of Maryland, College Park, MD 20742-0001, USA}

  \author[0000-0003-3841-9977]{Jian-Yang Li (李荐扬)}
  \affil{Planetary Science Institute, 1700 E. Ft. Lowell Rd., Tucson, AZ 85719, USA}

  \author[0000-0002-2668-7248]{Dennis Bodewits}
  \affil{Physics Department, Leach Science Center, Auburn University, Auburn, AL 36832, USA}

  \author[0000-0001-9328-2905]{Colin Snodgrass}
  \affil{Institute for Astronomy, University of Edinburgh, Royal Observatory, Edinburgh EH9 3HJ, UK}

  \author{Johannes Allen}
  \affil{Physics Department, Leach Science Center, Auburn University, Auburn, AL 36832, USA}

  \nocollaboration{100}

  \author[0000-0001-8018-5348]{Eric C. Bellm}
  \affiliation{DIRAC Institute, Department of Astronomy, University of Washington, 3910 15th Avenue NE, Seattle, WA 98195, USA}

  \author[0000-0002-8262-2924]{Michael W. Coughlin}
  \affil{School of Physics and Astronomy, University of Minnesota, Minneapolis, Minnesota 55455, USA}

  \author{Andrew J. Drake}
  \affiliation{Department of Astronomy, California Institute of Technology, 1200 E. California Blvd, Pasadena, CA 91125, USA}

  \author[0000-0001-5060-8733]{Dmitry A. Duev}
  \affiliation{Division of Physics, Mathematics, and Astronomy, California Institute of Technology, Pasadena, CA 91125, USA}

  \author[0000-0002-3168-0139]{Matthew J. Graham}
  \affiliation{Division of Physics, Mathematics, and Astronomy, California Institute of Technology, Pasadena, CA 91125, USA}

  \author[0000-0002-6540-1484]{Thomas Kupfer}
  \affiliation{Texas Tech University, Department of Physics \& Astronomy, Box 41051, 79409, Lubbock, TX, USA}

  \author[0000-0002-8532-9395]{Frank J. Masci}
  \affiliation{IPAC, California Institute of Technology, 1200 E. California Blvd, Pasadena, CA 91125, USA}

  \author{Dan Reiley}
  \affiliation{Caltech Optical Observatories, California Institute of Technology, Pasadena, CA 91125, USA}

  \author{Richard Walters}
  \affiliation{Caltech Optical Observatories, California Institute of Technology, Pasadena, CA 91125, USA}

  \collaboration{100}{Zwicky Transient Facility Collaboration}

  \author[0000-0002-3202-0343]{M. Dominik}
  \affil{Centre for Exoplanet Science, SUPA, School of Physics \& Astronomy, University of St Andrews, North Haugh, St Andrews KY16 9SS, UK}

  \author{U. G. J{\o}rgensen}
  \affil{Centre for ExoLife Sciences (CELS), Niels Bohr Institute, {\O}ster Voldgade 5, 1350 Copenhagen, Denmark}

  \author{A. Andrews}
  \affil{Department of Physical Sciences, The Open University, Milton Keynes, MK7 6AA, UK}

  \author{N. Bach-M{\o}ller}
  \affil{Centre for ExoLife Sciences (CELS), Niels Bohr Institute, {\O}ster Voldgade 5, 1350 Copenhagen, Denmark}

  \author{V. Bozza}
  \affil{Dipartimento di Fisica "E.R. Caianiello", Universit{\`a} di Salerno, Via Giovanni Paolo II 132, 84084, Fisciano, Italy}
  \affil{Istituto Nazionale di Fisica Nucleare, Sezione di Napoli, Napoli, Italy}

  \author{M. J. Burgdorf}
  \affil{Universit{\"a}t Hamburg, Faculty of Mathematics, Informatics and Natural Sciences, Department of Earth Sciences, Meteorological Institute, Bundesstra\ss{}e 55, 20146 Hamburg, Germany}

  \author[0000-0002-3913-3746]{J. Campbell-White}
  \affil{SUPA, School of Science and Engineering, University of Dundee, Nethergate, Dundee DD1 4HN, UK}

  \author{S. Dib}
  \affil{Centre for ExoLife Sciences (CELS), Niels Bohr Institute, {\O}ster Voldgade 5, 1350 Copenhagen, Denmark}
  \affil{Max Planck Institute for Astronomy, K\"{o}nigstuhl 17, D-69117 Heidelberg, Germany}

  \author[0000-0002-3648-0507]{Y. I. Fujii}
  \affil{Department of Physics, Nagoya University, Furo-cho, Chikusa-ku, Nagoya, Aichi, 464-8602, Japan}
  \affil{Graduate School of Human and Environmental Studies, Kyoto University, Yoshida-Nihonmatsu, Sakyo, Kyoto 606-8501, Japan}

  \author{T. C. Hinse}
  \affil{Institute of Astronomy, Faculty of Physics, Astronomy and Informatics, Nicolaus Copernicus University, Grudziadzka 5, 87-100 Torun, Poland}
  \affil{Chungnam National University, Department of Astronomy \& Space Science, 34134 Daejeon, South Korea}

  \author[0000-0003-0961-5231]{M. Hundertmark}
  \affil{Astronomisches Rechen-Institut, Zentrum f{\"u}r Astronomie der Universit{\"a}t Heidelberg (ZAH), 69120 Heidelberg, Germany}

  \author{E. Khalouei}
  \affil{Department of Physics, Sharif University of Technology, PO Box 11155-9161 Tehran, Iran}

  \author{P. Longa-Pe{\~n}a}
  \affil{Centro de Astronom{\'{\i}}a (CITEVA), Universidad de Antofagasta, Av.\ Angamos 601, Antofagasta, Chile}

  \author[0000-0003-2935-7196]{M. Rabus}
  \affil{Departamento de Matem\'atica y F\'isica Aplicadas, Universidad Cat\'olica de la Sant\'isima Concepci\'on, Alonso de Rivera 2850, Concepci\'on, Chile}

  \author{S. Rahvar}
  \affil{Department of Physics, Sharif University of Technology, PO Box 11155-9161 Tehran, Iran}

  \author{S. Sajadian}
  \affil{Department of Physics, Isfahan University of Technology, Isfahan 84156-83111, Iran}

  \author[0000-0003-1310-8283]{J. Skottfelt}
  \affil{Department of Physical Sciences, The Open University, Milton Keynes, MK7 6AA, UK}

  \author{J. Southworth}
  \affil{Astrophysics Group, Keele University, Staffordshire, ST5 5BG, UK}

  \author[0000-0002-9024-4185]{J. Tregloan-Reed}
  \affil{Instituto de Investigaci\'on en Astronomia y Ciencias Planetarias, Universidad de Atacama, Copiap\'o, Atacama, Chile}

  \author{E. Unda-Sanzana}
  \affil{Centro de Astronom{\'{\i}}a (CITEVA), Universidad de Antofagasta, Av.\ Angamos 601, Antofagasta, Chile}

  \collaboration{100}{MiNDSTEp Collaboration}

  \begin{abstract}
    Cometary activity is a manifestation of sublimation-driven processes at the surface of nuclei.  However, cometary outbursts may arise from other processes that are not necessarily driven by volatiles.  In order to fully understand nuclear surfaces and their evolution, we must identify the causes of cometary outbursts.  In that context, we present a study of mini-outbursts of comet 46P/Wirtanen.  Six events are found in our long-term lightcurve of the comet around its perihelion passage in 2018.  The apparent strengths range from $-0.2$ to $-1.6$ mag in a 5\arcsec{} radius aperture, and correspond to dust masses between $\sim10^4$ to $10^6$ kg, but with large uncertainties due to the unknown grain size distributions.  However, the nominal mass estimates are the same order of magnitude as the mini-outbursts at comet 9P/Tempel~1 and 67P/Churyumov-Gerasimenko, events which were notably lacking at comet 103P/Hartley 2.  We compare the frequency of outbursts at the four comets, and suggest that the surface of 46P has large-scale ($\sim$10--100~m) roughness that is intermediate to that of 67P and 103P, if not similar to the latter.  The strength of the outbursts appear to be correlated with time since the last event, but a physical interpretation with respect to solar insolation is lacking.  We also examine \textit{Hubble Space Telescope} images taken about 2 days following a near-perihelion outburst.  No evidence for macroscopic ejecta was found in the image, with a limiting radius of about 2-m.
  \end{abstract}

  \section{Introduction} \label{sec:intro}
  Comet 46P/Wirtanen is a small Jupiter-family comet that has been considered as a potential spacecraft target.  The effective radius is 0.6~km \citep{lamy98-wirtanen, boehnhardt02-wirtanen}, making it one of the smallest periodic comets \citep{snodgrass11-nuclei}.  The comet made an historic flyby of Earth in 2018, passing just 0.0775~au away (1.16\timesten{7}~km) on 2018 December 16 (JPL Horizons orbital solution K181/21).  The geometry with respect to the Earth and Sun was exceptionally favorable, with long observing opportunities and a total apparent magnitude peaking near $V\sim5$~mag (IAU Minor Planet Center Database).

  In many respects, comet Wirtanen is considered a near-twin of comet 103P/Hartley 2.  They have similar orbits, dust and gas production rates, and nuclear radii \citep{ahearn95, ahearn11}.  As a consequence, both comets are considered to be hyperactive, i.e., their water production rates suggest a sublimating surface area comparable to the total nuclear surface area, whereas most comets have a ratio $\lesssim10$\% \citep{ahearn95}.  Comet Hartley 2 was a flyby target of the \di{} spacecraft \citep{ahearn11} and the subject of a large observational campaign in 2010 \citep{meech11-epoxi}.  Thus, the 2018 perihelion passage of comet Wirtanen presented an opportunity to apply the knowledge gained from the studies of comet Hartley~2 to comet Wirtanen and the broader comet population.

  One important difference between Wirtanen and Hartley 2 is the lack of cometary outbursts in the latter \citep{ahearn11}.  Cometary outbursts are brief increases in mass loss \citep{hughes90}, instigated by mechanical or thermophysical processes, such as cliff collapse \citep{pajola17}, avalanches \citep{steckloff16-dps-outbursts}, nuclear fragmentation \citep{boehnhardt04-comets2}, or structural failure and release of pressure from a sub-surface gas reservoir \citep{agarwal17-outburst}, charged by, e.g., water ice phase changes \citep{patashnick74-water, bar-nun90} or gas dissolution from a liquid \citep{miles16-thermal}.  Outbursts of many comets have been observed, e.g., comets Kohoutek 1973f, Bowell 1980b, 9P/Tempel 1, and 67P/Churyumov-Gerasimenko \citep{ahearn75, ahearn84-bowell, ahearn05, ahearn16-lpsc-outbursts}, but none have been confirmed for comet Hartley 2.  This result is in spite of the 2010 observational campaign, and near-continuous photometry from the \di{} spacecraft.  In contrast, clear outbursts of comet Wirtanen were observed in 1991, 2002, 2008, and 2018 \citep{yoshida13-46p-2002, kidger04, kidger08, kronk17, combi20-soho, farnham19-wirtanen-tess}.

  Dense, long-term photometric and spectroscopic coverage of comets is needed to advance our understanding of cometary activity \citep{ahearn17-comets}.  Present-day wide-field time-domain surveys, such as the Zwicky Transient Facility \citep[ZTF;][]{bellm19-ztf,graham19-ztf} and the Asteroid Terrestrial-impact Last Alert System \citep[ATLAS;][]{tonry18-atlas}, can partially address this challenge with broad-band photometric imaging at a near-daily cadence.  In this work, we present a long-term lightcurve of comet Wirtanen and examine it for evidence of outbursts in activity.  This paper is a follow-up to the preliminary investigation by \citet{kelley19-outbursts}.

  \section{Observations and Data}\label{sec:obs}
  Broad-band images of comet Wirtanen were obtained from four observatories in 2018 and 2019: Palomar Observatory, Lowell Observatory, the European Southern Observatory, and the \textit{Hubble Space Telescope}.  We first describe the ground-based data, which we use to form a long-term lightcurve of coma, then the \hst{} data, which were taken as part of a \textit{Chandra X-Ray Observatory} campaign to study charge exchange in the cometary coma \citep{bonamente20-this-issue}.

  \subsection{Ground-based Observatories}
  \subsubsection{Palomar Observatory}
  Observations of comet Wirtanen were identified in the ZTF Data Release 3, Partnership, and Caltech archives with the ZChecker program \citep{kelley19-zchecker}.  ZTF is a wide-field time-domain survey using the Samuel Oschin 1.2-m telescope at Palomar Mountain with a 16-CCD camera.  Each 6144$\times$6160 CCD has a 1\farcs01 pixel scale, yielding a total camera field of view of 47~deg$^2$ with an 86\% fill factor \citep{bellm19-ztf}.  The robotic system executes multiple simultaneous surveys, with a range of science goals \citep{graham19-ztf}.  Comet Wirtanen was found in 352 images in total ($g$, $r$, and $i$ bands, 30-s exposure times), taken between 2018 July 13 and 2019 June 06 UTC (84 nights), observed in the Northern Sky, Galactic Plane, Asteroid Rotation, $i$-band, and One-Day Cadence surveys \citep{bellm19-surveys}.  Most nights have only one or two images, except during the Asteroid Rotation survey, which observed 46P over 3- to 4-hour periods on 2019 January 24, 25, and 26 UTC with a 255-s cadence.  All data were reduced with the ZTF data pipeline \citep{masci19-ztf}.  The processing typically includes reference image subtraction, which removes smooth background and photometrically stable celestial objects, leaving image artifacts and transients (including solar system objects).  We find no significant difference between small-aperture ($<10$~pix) photometry measured with or without the reference subtracted data, except that the latter are less likely to be affected by background stars.  Therefore, we use reference subtracted data whenever possible for photometry.  When the comet is bright and the angular extent is large, the morphology is best studied without reference subtraction.

  \subsubsection{Lowell Observatory}
  Images of comet Wirtanen were taken with the Lowell Observatory 0.8-m robotic telescope located at Anderson Mesa \citep{buie10-robo31} through an $R$-band filter between 2018 September 23 and 2019 February 08 UTC (26 nights).  The camera uses a 2048$\times$2048 CCD with a pixel scale of 0\farcs45, yielding a 15\arcmin{} field of view.  Standard image bias and flat-field corrections were applied.  Typically 3 images were taken per night, with 12- to 300-s exposure times and the telescope tracking at the rate of the comet.

  \subsubsection{European Southern Observatory, La Silla}
  After combining the ZTF and Lowell data sets, we identified a gap in temporal coverage in early August.  Select images taken with the Danish 1.54-m telescope at La Silla Observatory were reduced and examined in order to fill this gap.  Observations utilized the Danish Faint Object Spectrograph and Camera (DFOSC), which has a field of view of 13\farcm7$\times$13\farcm7 and a pixel scale of 0\farcs39, and were taken on an approximately weekly cadence between 2018 June 18 and September 17 UTC (8 nights), primarily in the $R$-band. Additional images were taken in $UBVRI$-bands later in this period but are not included in the work presented here.

  \subsubsection{Photometry}
  All ground-based data are calibrated to the PS1 photometric system using background stars in each field.  The calibration of the ZTF data are described by \citet{masci19-ztf}.  The remaining data were calibrated to the $r_{\mathrm{P1}}$-band (i.e., PS1 system) using the ATLAS Refcat2 photometric catalog \citep{tonry18-refcat2} and Calviacat software \citep{kelley19-calviacat}.  Uncertainties in the absolute calibrations are propagated into the final measurement errors, but a minimum uncertainty of 0.02~mag is assumed.  All data are color corrected using the measured coma colors (Section \ref{sec:results}) and photometric calibration solutions.  Photometry within a constant angular aperture radius of 5\arcsec{} is given in Table~\ref{tab:obs}, with 372 data points taken on 111 unique nights spanning 352 days.

  Although the comet is bright, it does not saturate the ZTF detectors.  In 30-s exposures, the saturation limit for point sources is about 13~mag, depending on the filter.  Since the comet is an extended source, and our photometry is in a 5\arcsec{} radius aperture (whereas seeing is typically around 2\arcsec{} FWHM), the comet data are not saturated despite the bright photometric values reported in this work ($r\gtrsim11$~mag).

  \begin{deluxetable*}{cccl}
    \tablecaption{Comet 46P/Wirtanen geometric, photometric, and derived data. \label{tab:obs}}
    \tablehead{
      \colhead{Column}
      & \colhead{Name}
      & \colhead{Unit or scale}
      & \colhead{Description}
    }
    \startdata
    (1) & Source & \nodata & Name of telescope \\
    (2) & Date & UTC & Mean time of observations \\
    (3) & $T-T_P$ & days & Time offset from perihelion\tablenotemark{a} \\
    (4) & \rh & au & Comet heliocentric distance \\
    (5) & $\Delta$ & au & Comet-observer distance \\
    (6) & $\theta$ & deg & Sun-comet-observer (phase) angle \\
    (7) & Filter & \nodata & Filter name \\
    (8) & Exposure & s & Total exposure time \\
    (9) & Airmass & \nodata & Mean airmass of observations \\
    (10) & Seeing & arcsec & FWHM of (potentially trailed) point sources \\
    (11) & $m$ & mag & Apparent magnitude in 5\arcsec{} radius aperture (PS1 system) \\
    (12) & $\sigma_m$ & mag & Uncertainty on $m$ \\
    (13) & Trail & mag & Trailed-source correction applied to ZTF photometry \\
    (14) & Trend & mag & $r$-band magnitude trend from piecewise fit \\
    (15) & \afrho[\theta] & cm & Comet photometric quantity, based on $m$ \\
    (16) & $G$ & km$^2$ & Geometric cross-section, based on $m$
    \enddata
    \tablenotetext{a}{$T_P=$2018 December 12.94146 UTC \citep{mpc114607}.}
    \tablecomments{Table 1 is published in its entirety in the machine-readable format.
      The column descriptions are shown here for guidance regarding its content.}
  \end{deluxetable*}

  In contrast with the Lowell and Danish telescope observations, the ZTF survey data images are tracked in the Celestial reference frame, causing the comet to trail during the 30-s exposures.  With non-sidereal rates up to $\sim$600\arcsec{}~\inv{hr}, the comet tailed 0.5--6\arcsec{} per exposure.  Thus, photometry in a 5\arcsec{} radius aperture may be affected.  We attempt to correct for those losses by generating an image of an idealized coma (surface brightness proportional to $\rho^{-1}$, where $\rho$ is the projected distance to the nucleus) and convolving it with a linear kernel.  The length of the kernel is equal to the calculated trailed length per exposure, and the correction factor is the ratio of the brightness of the trailed coma to that of the ideal coma, measured in a 5\arcsec{} radius aperture.  The corrections range from $-0.01$ to $-0.11$~mag (Table~\ref{tab:obs}), and are applied to all ZTF photometry.  Assuming a shallower profile, e.g., $\rho^{-0.8}$, affects the correction by $\leq0.02$~mag.

  \subsection{Hubble Space Telescope}
  \textit{Hubble Space Telescope} (\hst) imaged comet Wirtanen with the Wide Field Camera 3 (WFC3) UVIS channel at two epochs.  Each epoch contained four \hst{} orbits, organized into two two-orbit groups separated by one orbit, covering about 7 hours in duration.  The data spanned 2018 December 13 09:15 to 16:18, and December 25 10:30 to 17:33 UTC.  The comet was observed through two mid-band filters F689M and F845M (11\% wide bandpass) with the 2k$\times$2k sub-frame, which has a field of view of 80\arcsec{}$\times$80\arcsec{} given the 0\farcs04 pixel scale.  Due to the non-linear non-sidereal movement of the comet and the high spatial resolution of the WFC3/UVIS camera, the comet was trailed by up to 4 pixels for all F689M images except one with an 8-pixel trail, and by various amounts up to 9 pixels in the F845M images, despite the short exposure times of 10 and 16~s used for F689M and F845M, respectively.  On the other hand, all images are well exposed with the peak brightness up to 24\% of the saturation level.

  Photometric measurements are based on the images reduced by the standard WFC3 calibration pipeline \citep{gennaro18}.  To remove cosmic rays, we divided each image into a grid of 20$\times$20-pixel boxes, then clipped and replaced 3$\sigma$ outliers with the mean in each box.  The center 40$\times$40 pixel region was excluded from this cosmic ray removal process in order to preserve the inner coma.  For the fragment search, we also removed cosmic rays with the LA Cosmic algorithm \citep{vandokkum01-lacosmic}.  Sky background was estimated by the mean of four 100$\times$100 pixel boxes near the corners of the images.  The pixel area map of the corresponding detector chip was applied to correct for pixel area change in the spatially distorted (FLT) images before photometric measurement.  The total count was then measured in a 5\arcsec-radius aperture and converted to flux and apparent magnitude following the photometric calibration constants \citep{gennaro18}.  Our photometry is limited by the absolute photometric uncertainty for WFC3/UVIS images (2\%).  The effect of source trailing in our images is negligible for 5\arcsec{}-radius aperture photometry.

  \section{Results} \label{sec:results}
  \subsection{Coma Color}
\begin{figure}
    \plotone{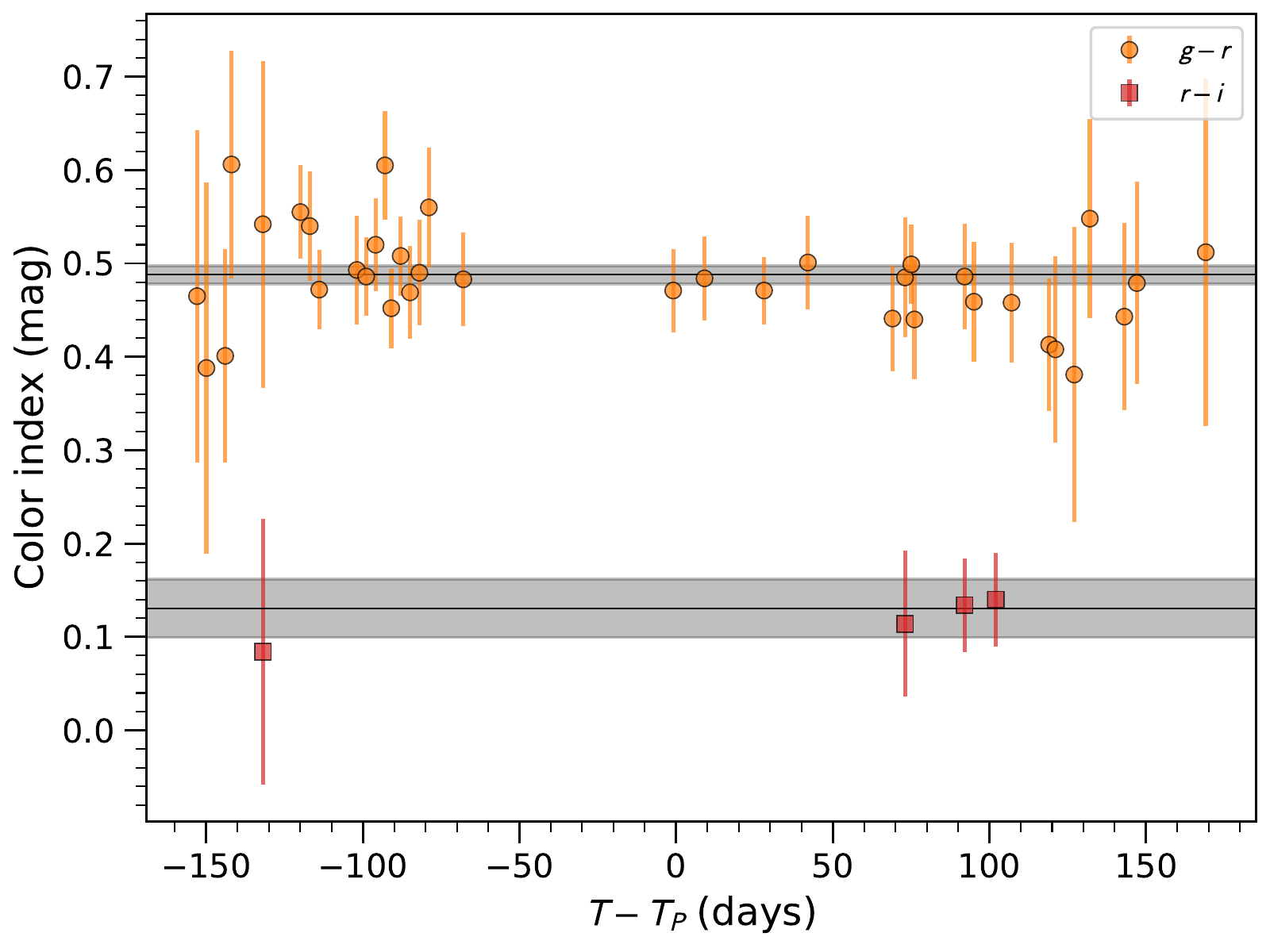}
    \caption{Color index versus time from perihelion ($T-T_P$) for comet 46P/Wirtanen measured with Zwicky Transient Facility photometry in the $g$, $r$, and $i$ bands.  The means and uncertainties are drawn as horizontal lines and shaded areas, respectively.}\label{fig:color}
  \end{figure}

\begin{figure*}
    \plotone{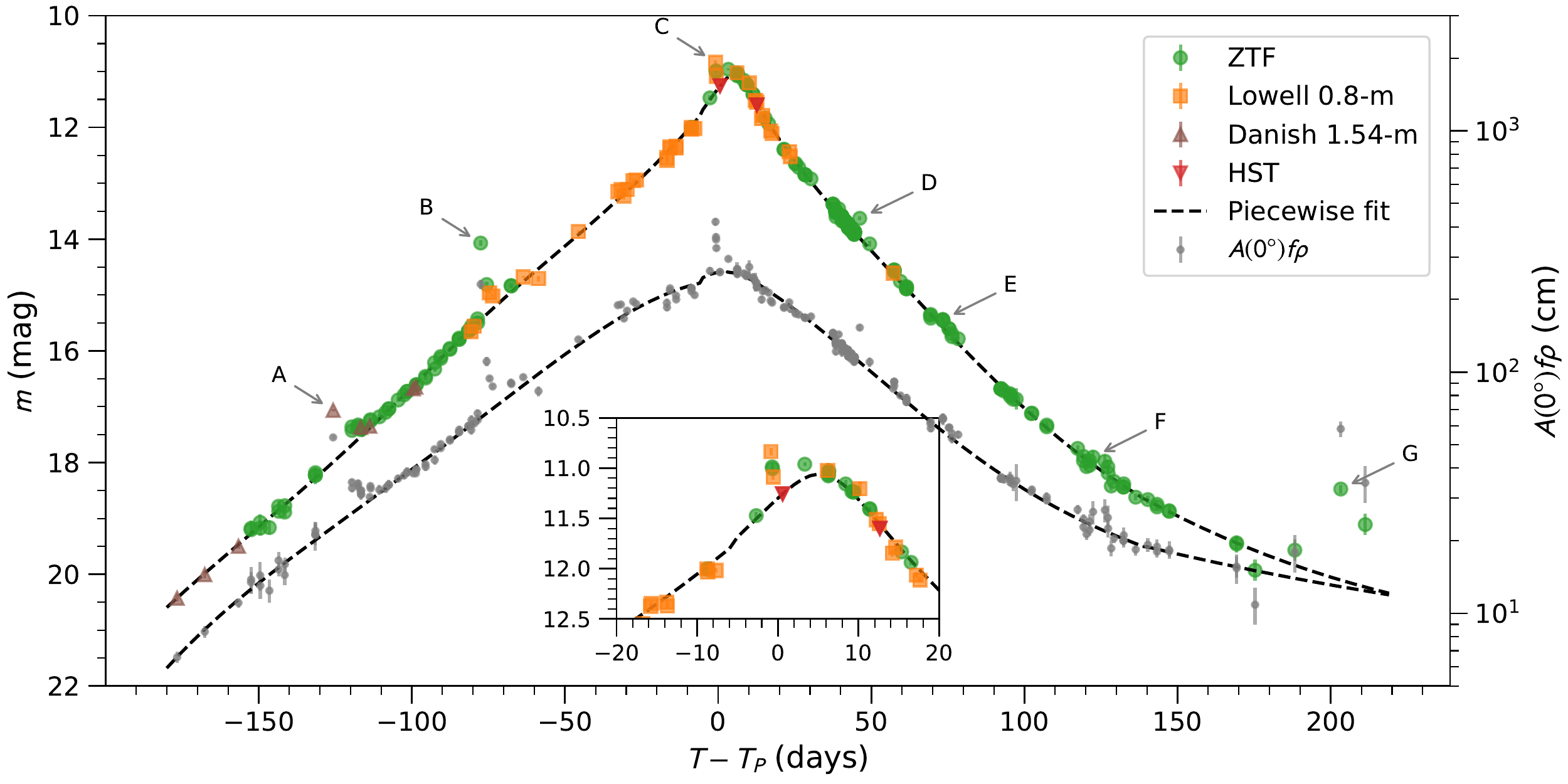}
    \caption{Lightcurve of comet 46P/Wirtanen measured within 5\arcsec{} radius apertures.  Photometry from the $g$- and $i$-band have been scaled with the measured coma colors to make an effective $r$-band data set.  Also shown is the photometry converted to the \afrho[0\degr] quantity.  A trend-line based on a piece-wise fit to the photometry is shown as a dashed line (see Section~\ref{sec:results} for details).  Seven sets of anomalous data points are labeled A--G.}
    \label{fig:lightcurve}
  \end{figure*}

  The $g-r$ color of comet Wirtanen was previously measured from a limited set of ZTF photometry by \citet{kelley19-outbursts} to be 0.45$\pm$0.02~mag.  We compute $g-r=0.49\pm0.01$~mag and $r-i=0.13\pm0.03$~mag from the weighted means of 36 and 4 nightly color measurements, respectively.  Those colors appear to be consistent throughout the data set (Fig.~\ref{fig:color}), with the largest deviation at the $2.0\sigma$ level (reduced $\chi^2$ is 0.5 for $g-r$, 0.1 for $r-i$).  The mean color from \hst{} is $m_{689}-m_{845}=0.15\pm0.02$~mag.  To convert the \hst{} photometry into $r$-band data, we use the measured \hst{} color, and extrapolate it to the PS1 $r$-band with a spectrum of the Sun.  Throughout this work, we adopt the composite spectrum of the Sun from \citet{haberreiter17-sun} and \citet{willmer18-sun} for filter calibrations (we estimate the apparent magnitude of the Sun in the F689M and F845M filters to be $-27.01$ and $-27.07$~mag, AB magnitude system).  Based on the \hst{} color, we calculate $r-m_{689}=0.04$~mag.  Using these colors, an effective $r$-band lightcurve versus time from perihelion is shown in Fig.~\ref{fig:lightcurve}.

  The colors of the coma correspond to spectral gradients \citep{ahearn84-bowell} of \specgrad{g,r}{6.8\pm0.7} \specgrad{r,i}{0.7\pm2.0}, and \specgrad{689\mathrm{M},845\mathrm{M}}{5.2\pm1.2}, where the subscripts denote the bandpasses used in the calculations.  The $S_{r,i}$ and $S_{689\mathrm{M},845\mathrm{M}}$ are consistent at the $2\sigma$ level.  Note that these colors are not necessarily that of the dust coma, as there are gas emission bands present at these wavelengths, especially \ch{C2} in $g$, but also \ch{NH2} in $r$, $R$, and F689M, and CN in $i$.  For example, \citet{zheltobryukhov20-wirtanen} estimate a gas contamination of 5\% in the $V$-band in a 1\arcsec{} (100~km) radius aperture, a fraction that should grow with aperture size given the different radial profiles of C$_2$ and dust.  See \citet{fink98-wirtanen} and Fig.~1 of \citet{ponomarenko18-wirtanen} for figures showing relevant optical spectra of comet Wirtanen.

  We searched the literature for other comet Wirtanen coma colors and compared them to our values, finding reasonable agreements.  A spectrum of the comet by \citet{ponomarenko18-wirtanen} has \specgrad{}{8.6} over the wavelength range 480--750~nm (no uncertainties were quoted).  \citet{lamy98-wirtanen} measured the coma to have \specgrad{V,R}{8.3\pm7.7} in \hst{} filter photometry.  \citet{zheltobryukhov20-wirtanen} measured neutral-to-blue colors in $BVRI$ broadband photometry on 2019 February 8 and 10 UTC ($T-T_P$=57--60~days) in a 5000~km aperture radius (compare with our 5\arcsec=1330~km radius).  In terms of spectral slope, they report \specgrad{V,R}{-16.7\pm7.1} and $-7.5\pm16.3$, \specgrad{R,I}{-8.7\pm4.8} and $-8.1\pm8.4$.  Despite the nominally blue spectral slopes, the uncertainties are large enough to be in agreement with our estimates but at 2 to 3$\sigma$ level for the better quality measurements.  Their observations fall in a gap in our ZTF color coverage (Fig.~\ref{fig:color}, $T-T_P=57-60$~days), but we can make an estimate on February 8 by comparing Lowell 0.8-m photometry to ZTF photometry, and find $g-r=0.45\pm0.03$~mag, which is in agreement with our average color (1.3$\sigma$ difference).  In addition, the photometric coverage is good starting February 8, and we find no unusual activity at this time.

  \subsection{Quiescent Activity}\label{sec:quiescent}
  In order to identify outbursts, it helps to define the quiescent activity trend.  We use the \afr{} model of \citet{ahearn84-bowell}.  This quantity is based on the brightness of the coma within a circular aperture.  Formally, it is the product of grain albedo ($A$), filling factor within the photometric aperture ($f$), and aperture radius ($\rho$, projected length at the distance of the comet).  \afr{} carries the units of $\rho$, but is proportional to dust mass-loss rate under idealized assumptions, e.g., a coma in free expansion with a constant production rate, grain size distribution, and composition (i.e., 1/$\rho$ surface brightness profile), and photometry free of gas contamination.  (See \citet{fink12} for more discussion on the physical interpretation.)  The albedo is commonly expressed as a function of phase angle, $\theta$, in order to explicitly account for the phase effect from non-isotropic scattering of sunlight by coma dust.  For the phase correction, we adopt the Schleicher-Marcus phase function, first used by \citet{schleicher11}.  In Table~\ref{tab:obs}, all photometry is converted to \afrho[0\degr].  In Fig.~\ref{fig:lightcurve}, we plot the effective $r$-band \afrho[0\degr] values after accounting for the measured color differences.

  We fit the $\log\afrho[0\degr]$ data with a polynomial as a function of either $\log\rh$ or time.  Candidate outbursts were excluded from the fit.  The best fit to the entire lightcurve is $(263\pm1)\ \rh^{-4.01\pm0.01}$~cm (RMS 0.04~mag).  However, we found this trend does not have sufficient precision for quantifying outbursts, with local deviations as strong as 22\%.  Therefore, we split the lightcurve into three segments with break points based on time from perihelion, $T-T_P=-5$ and +15~days.  Each segment is fit with 3rd or 4th degree polynomials versus time.  The RMS of the residuals are 0.07, 0.05, and 0.04~mag (excluding possible outbursts).  We will show that an outburst occured at the end of our lightcurve coverage.  The polynomial fit cannot be used to extrapolate the pre-ouburst quiescent lightcurve to the epochs of the outburst.  Therefore, for photometry after 150 days, we use a power-law extrapolation based on \rh, with a best-fit slope of $-1.67\pm0.44$ fit to the data at $T-T_p=$130 to 202 days.  We plot the piecewise trend in Fig.~\ref{fig:lightcurve} and report the trend values for each observation in Table~\ref{tab:obs}.

  The piecewise approach handles the near-perihelion photometry separately from the rest of the data, and allows for short- and long-term asymmetries around perihelion.  Near perihelion, the geometrical circumstances vary rapidly.  The comet moves 70\degr{} on the sky and through opposition, which occurred 6~days after perihelion.  Thus, the projection of the potentially non-isotropic coma onto the sky changes substantially, which affects the small aperture photometry.  We find that the \afrho[0\degr] is near constant from $-3$ to $+9$~days (Table~\ref{tab:obs}), aside from an outburst at $-1$~day and a single-point outlier on day +3 (Fig.~\ref{fig:lightcurve}, inset).  Moreover, the near-perihelion \afrho[0\degr] values are elevated by about 20\% with respect to the adjacent pre- and post-perihelion trends.

\begin{figure*}
    \plotone{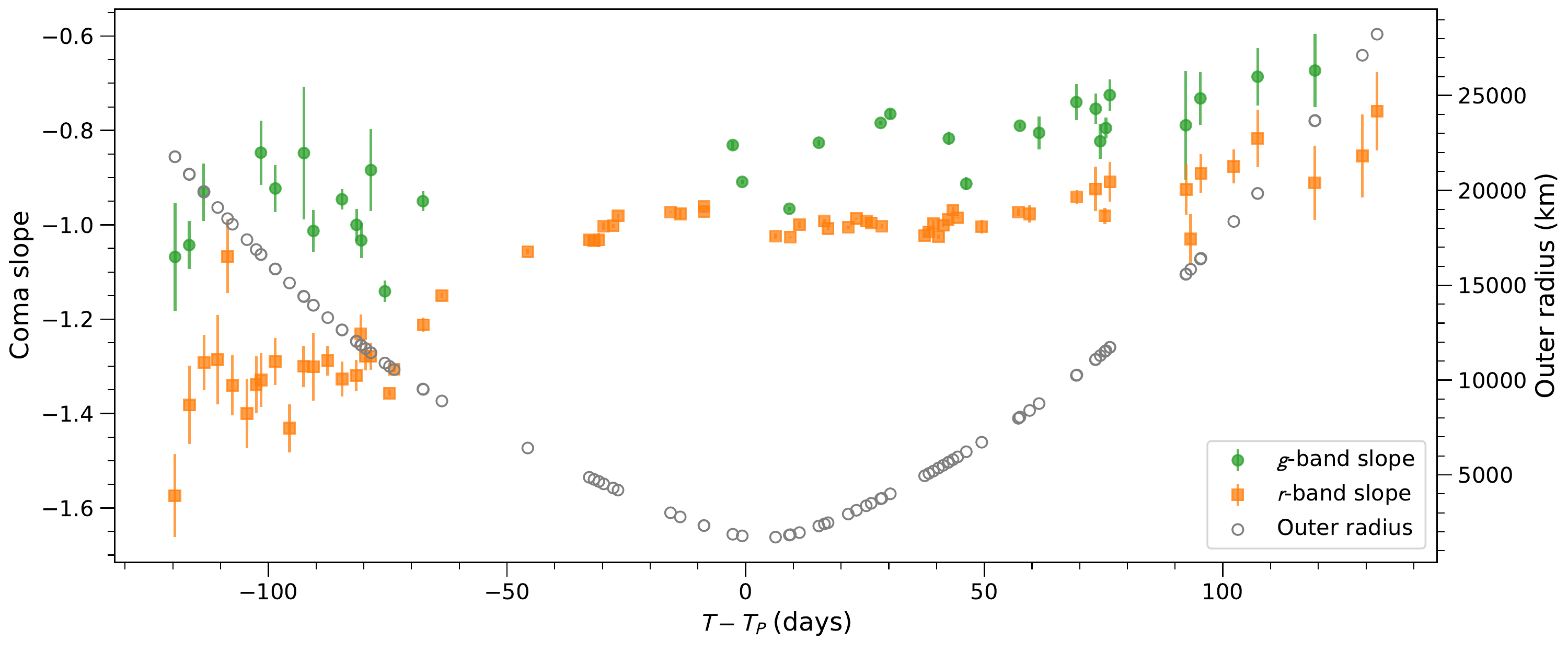}
    \caption{Azimuthally averaged radial profiles for $\rho<30$\arcsec{} versus time with respect to perihelion, based on ZTF and Lowell 0.8-m data.  The $g$- and $r$-band are shown separately.  Also plotted are the outer radii of the fit regions as projected lengths at the distance of the comet.}
    \label{fig:slope}
  \end{figure*}

  Note that our best-fit trends depend on the idealized assumptions of the \afr{} model (especially the assumption of a 1/$\rho$ surface brightness profile), our adopted phase curve, and our photometric aperture size (280--7600~km).  The goal of our investigation is to identify and characterize outbursts in the comet's activity, and the piecewise best-fit trend will serve this purpose, but may not be appropriate for other contexts.  \explain{Moved next sentence up two paragraphs.}  To aid in the interpretation of the trends, we fit the azimuthally averaged radial profiles at $\rho<30$\arcsec{} for the ZTF and Lowell 0.8-m images and plot the results in Fig.~\ref{fig:slope}.  Fits with a reduced-$\chi^2$ statistic $>2$, e.g., due to nearby stars or outburst ejecta, were ignored.  The $g$- and $r$-band data are separately fitted.  The $g$-band profiles are shallower than the $r$-band profiles: minimum/median/maximum = --1.1/--0.8/--0.7 for $g$, --1.6/--1.0/--0.8 for $r$.  This difference is consistent with the expectation that the $g$-band data includes emission from \ch{C2} gas, which has a surface brightness distribution shallower than 1/$\rho$ for these length scales \citep{combi04}.   Within the 30\arcsec{} radius, the $r$-band data transitions from tail-dominated ($\sim\rho^{-1.4}$) to coma-dominated ($\sim\rho^{-1}$) by $T-T_P=-30$~days.  The $r$-band coma remains near $\rho^{-1}$ for  $-30<T-T_P<60$~days, i.e., inside a radius of 5000--8000~km, after which it becomes slightly shallower, finishing near $\rho^{-0.85}$.  The asymmetry in slopes about perihelion may be due to slow moving grains, lingering near the nucleus.  This interpretation is consistent with the \afrho{} asymmetry, which is higher post-perihelion, and requires dust grains moving at \mps{} speeds ($2\times10^4$~km / 130~days since perihelion).

  \subsection{Outbursts}\label{sec:morphology}
\begin{figure*}
    \plotone{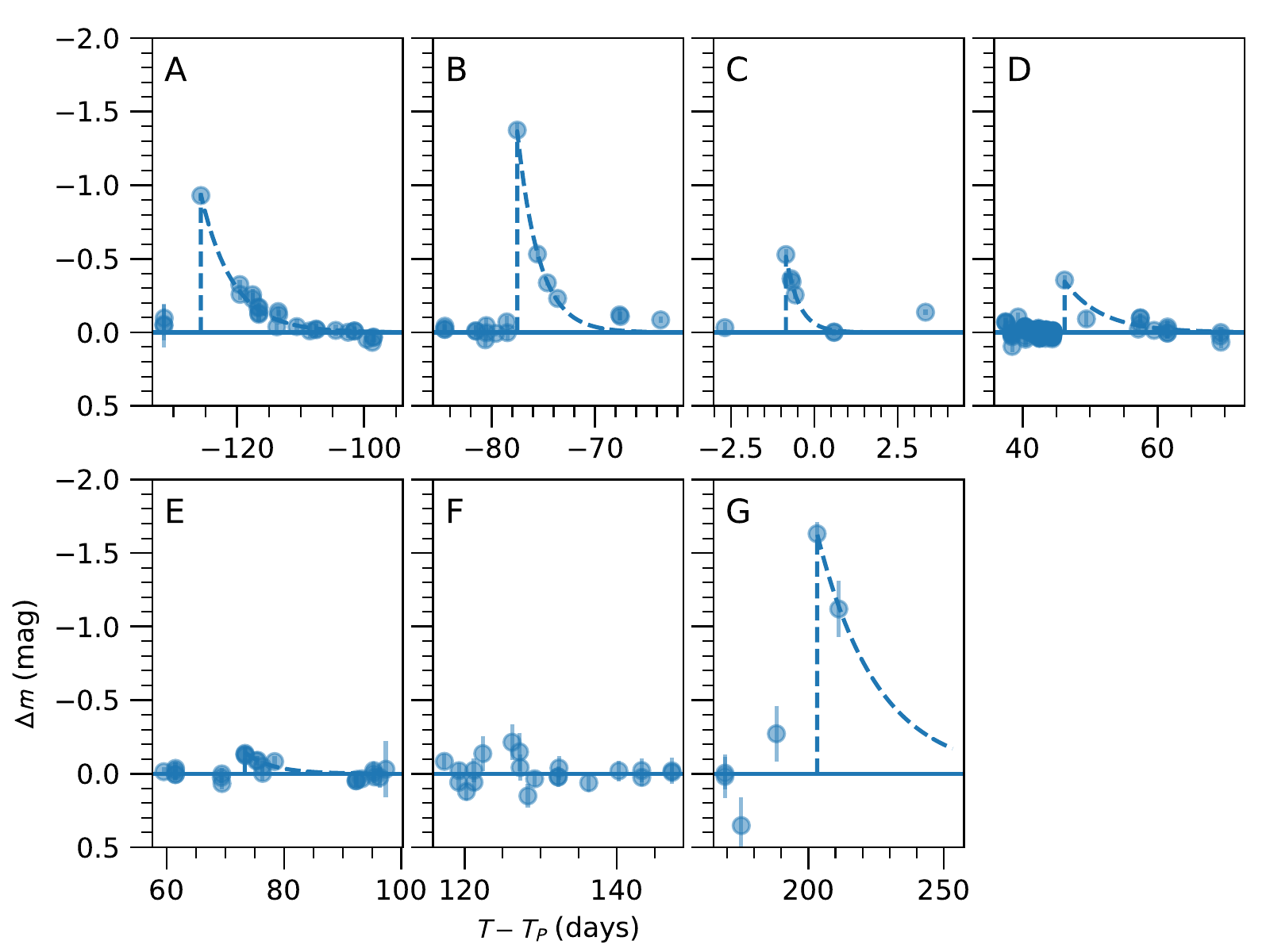}
    \caption{Lightcurves of six anomalous sets of data points, labeled A--G, identified in the lightcurve of 46P/Wirtanen (Fig.~\ref{fig:lightcurve}).  For each set, the baseline photometric trend has been removed, and an illustrative exponential function is shown as a dashed line.  Events A--C and G appear to be outbursts, characterized by a rapid brightening and exponential fading.  Event D is sparsely observed, but confirmed as an outburst by image morphology.  Event E appears to be a real deviation from the trend, but is not obviously an outburst.  Event F was not confirmed in the image morphology (Section \ref{sec:morphology}).}
    \label{fig:outbursts}
  \end{figure*}

  From manual inspection of the lightcurve, we identify seven sets of significant photometric outliers, labeled A--G in Fig.~\ref{fig:lightcurve}.  Figure~\ref{fig:outbursts} shows each set of data, after removing the photometric trends.  All but event F appear to be brightening events (i.e., outbursts).  Event C is followed by a single-point outlier 4 days later on 2018 December 16 UTC (Fig.~\ref{fig:lightcurve}, inset).  The rapid changes about perihelion,  and the fact that the photometry is sparse around this point (it is the only data between December 13 and 19), makes defining the quiescent activity at that time more challenging, therefore we do not interpret this point as an outburst.  Event F is also difficult to interpret, due to the weak peak brightness ($\sim-0.2$~mag), and a possible change in the quiescent trend at the same time.  Therefore, we only report F as a possible anomaly.

\begin{figure*}
    \plotone{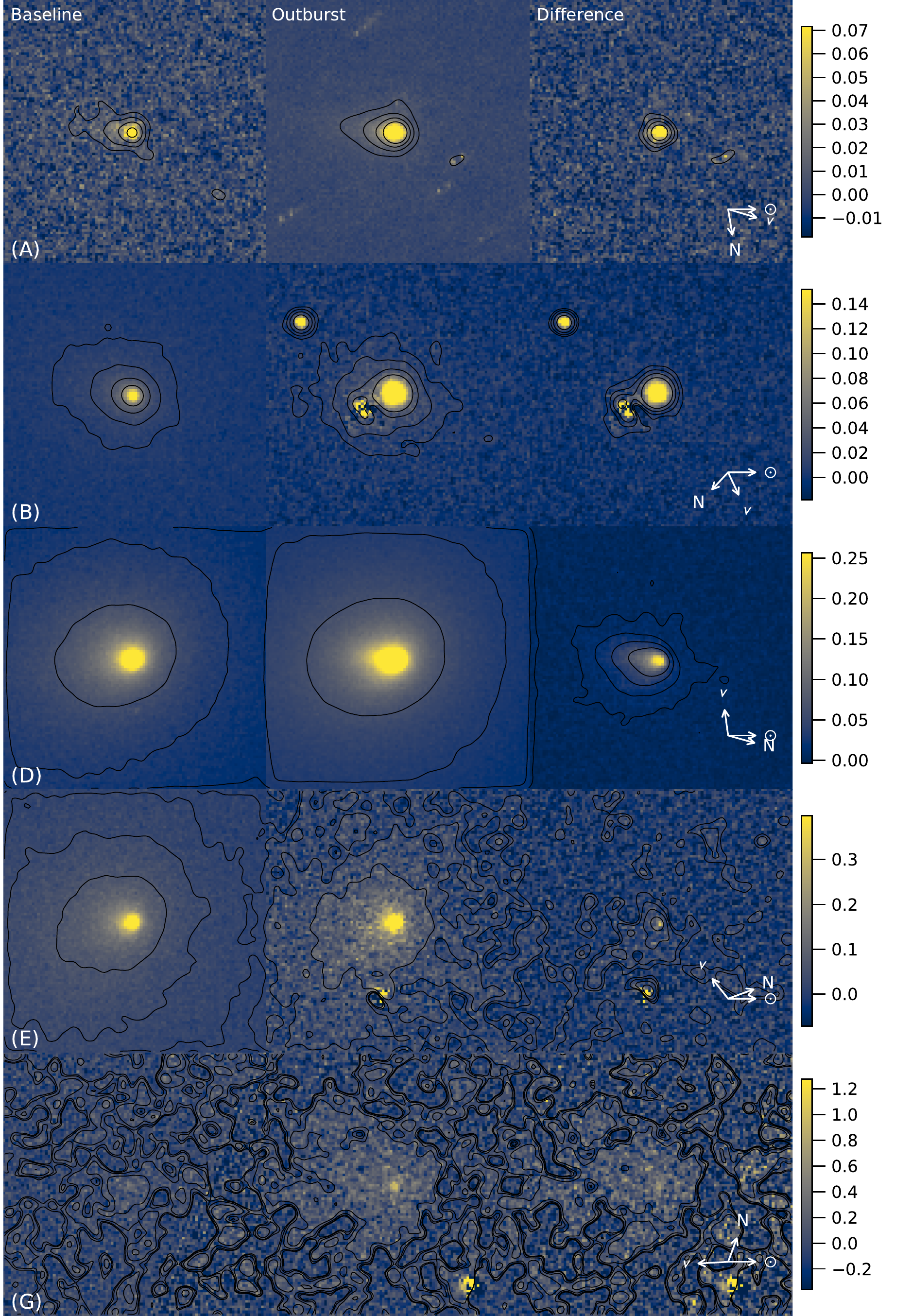}
    \caption{(Left and center) Baseline and outburst images for events A, B, D, E, and G based on ZTF and Danish 1.54-m data.  (Right)  Difference between outburst and the scaled baseline data.  All images are 1.7\arcmin$\times$1.7\arcmin, and scaled with respect to the peak of the comet in the outburst image as indicated by the colorbar.  Smoothed contours are spaced at factors of two intervals, the brightest of which is at 6.25\% of the peak.  The projected comet-Sun ($\odot$), comet velocity ($v$), and Celestial north (N) vectors are shown for the outburst image.  For outburst B, the artifacts near the comet are residuals after removing a nearby source.}
    \label{fig:abde}
  \end{figure*}

\begin{figure*}
    \plotone{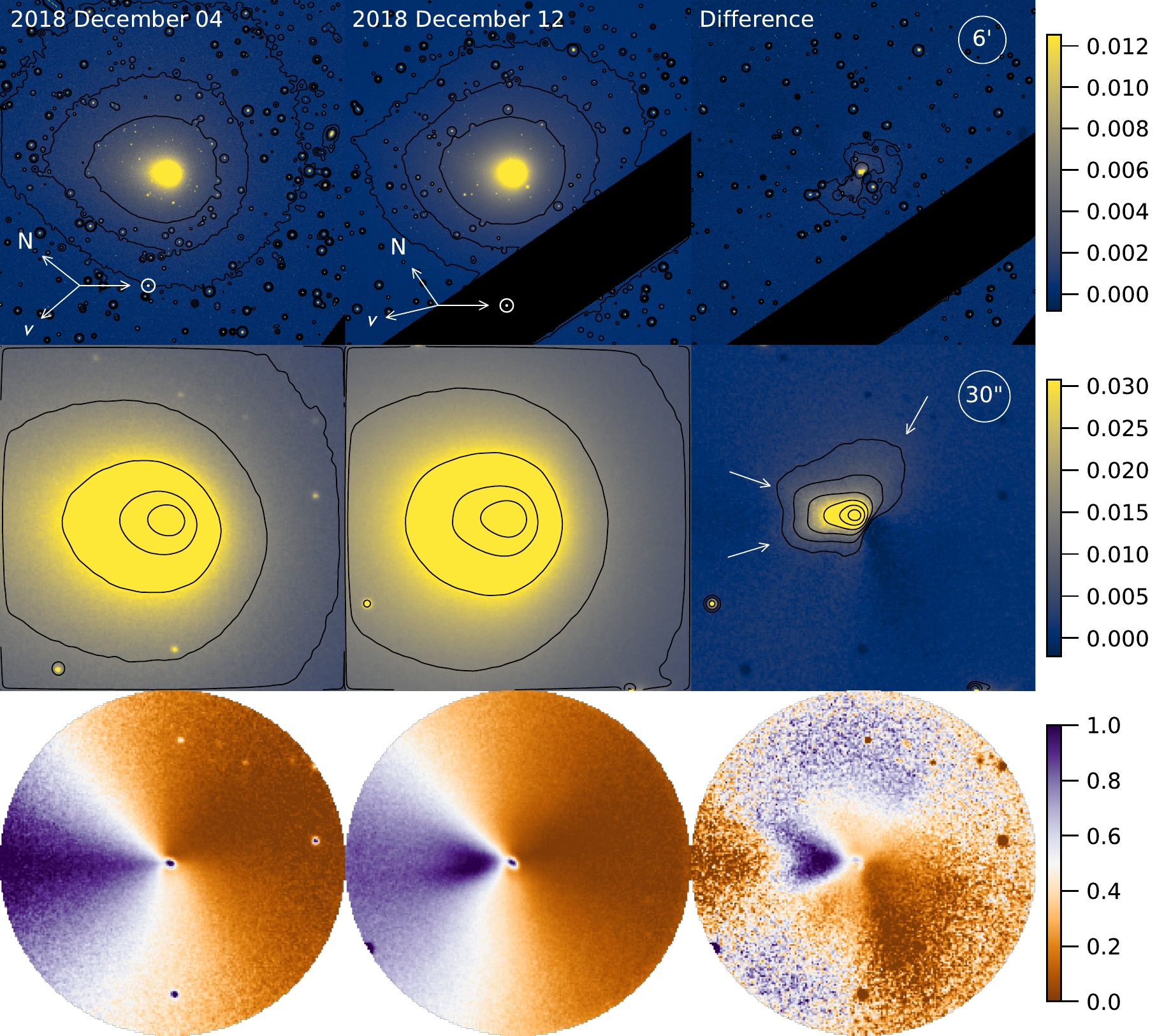}
    \caption{Same as Fig.~\ref{fig:abde} but for event C.  (Top) 44\arcmin{}$\times$44\arcmin{} field of view with smoothed contours spaced at factors of two intervals, the brightest of which is at 0.20\% of the peak.  (Center) 3\farcm4$\times$3\farcm4 field of view, the brightest contour is 12.5\% of the peak.  The masked region is a gap between the CCDs.  Arrows mark three outburst features.  (Bottom) Same as the center, but enhanced by normalizing the data with the azimuthal average, and displayed on a linear scale from the coma minimum to maximum.  Note the change in morphology after the outburst with the addition of a v-shaped pattern in the anti-sunward direction.  Projected vectors are provided for the baseline and outburst images.}
    \label{fig:c-ztf}
  \end{figure*}

\begin{figure*}
    \plotone{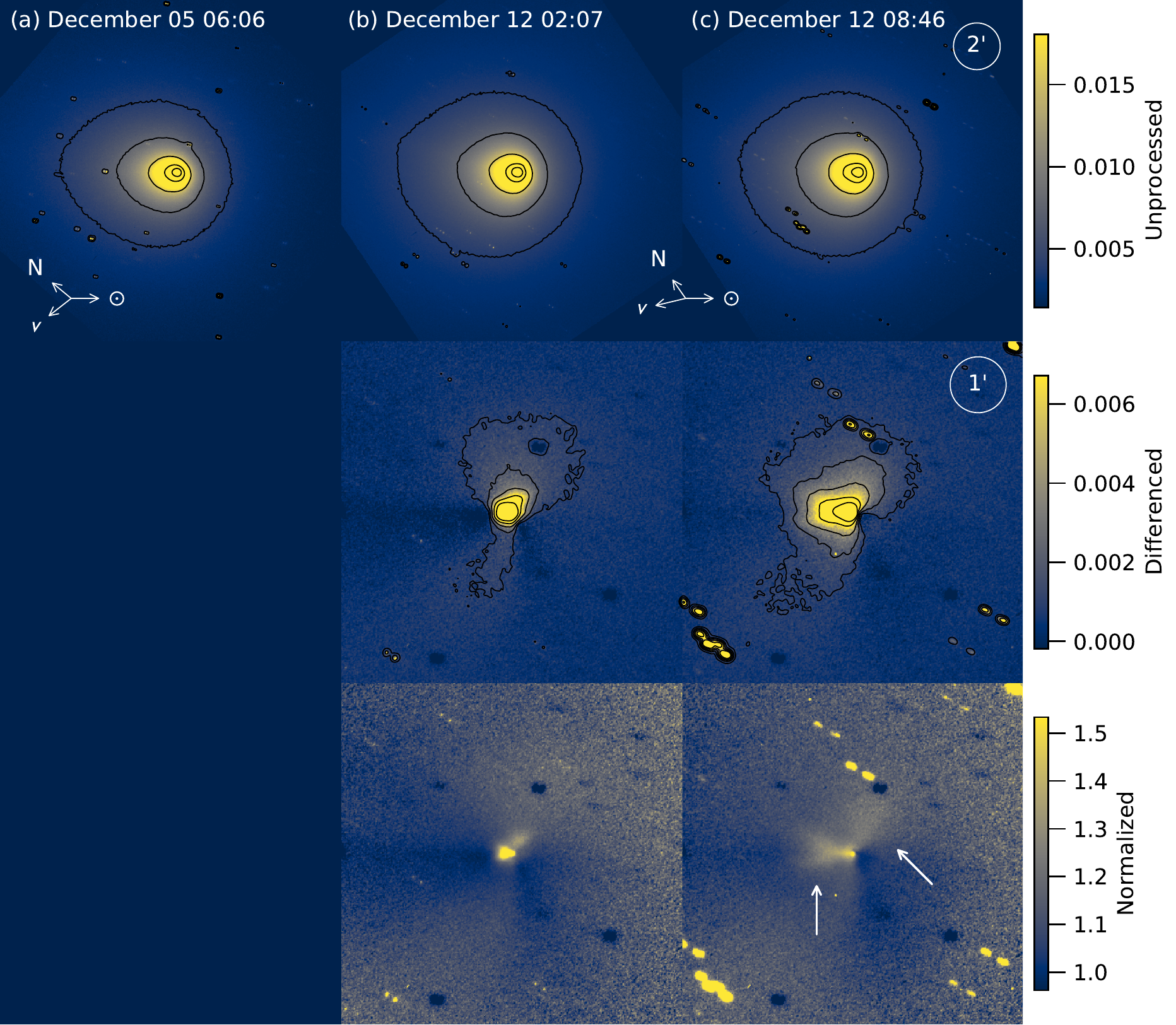}
    \caption{Lowell 0.8-m images taken (a) 2018 December 05, (b) 2018 December 12 at 02:07, and (c) 2018 December 12 at 08:46 UTC.  (Top) Unprocessed images and smoothed contours.  (Center) Image residuals after subtracting image (a), photometrically scaled according to our baseline photometric trend.  (Bottom) Images normalized by the scaled image (a).  Arrows indicate two prominent ejecta features in image (c).}
    \label{fig:c-lowell}
  \end{figure*}

  We visually inspected the candidate outburst image sets for supporting morphological evidence.  Because the unresolved nucleus is the ultimate source of any ejecta, the morphology of an outburst is initially that of a point-source, until the ejecta has moved far enough from the nucleus to be detectable as an extended source (as image sensitivity allows).  For each event, we defined one or more pre-event images to be used as a baseline model that was scaled and subtracted from the post-event data.  By inspection of the residuals, we can help identify the cause of the photometric anomalies.  The data were processed with the IPAC Montage software \citep{jacob10-montage} to scale images to a common pixel scale, place the comet at the center of the field, and align the projected Sun direction along the +$x$-axis.  The images are photometrically scaled to the post-event circumstances using the best-fit lightcurve trend, then median combined and subtracted from a post-event image to reveal the putative outburst ejecta.  Events A, B, D, E, and G are shown in Fig.~\ref{fig:abde}, and event C in Figs.~\ref{fig:c-ztf} and \ref{fig:c-lowell}.  Details on all sets follow.  Comments on the ejecta distributions are based on visual inspection of the images and radial profiles; position angles are measured eastward of Celestial north.  Photometry of the residuals are reported in Table~\ref{tab:outbursts}.
  \begin{itemize}
    \item[(A)] Seven ZTF images taken from 2018 July 22 to 2018 August 03 UTC were combined and subtracted from the three median combined Danish 1.54-m $R$-band images taken on August 09.  The residuals are extended, but still centrally peaked at the nucleus, and wholly contained within a 7\farcs1-radius aperture.

    \item[(B)] Six ZTF images taken from 2018 September 22 to 25 UTC were combined and subtracted from the ZTF $i$-band image taken September 26.  The ejecta is nearly point-source like, but slightly extended towards PA$\sim$270\degr.  This direction is inconsistent with the proper motion trailing, which is 0\farcs4 along PA=295\degr.  A nearby star limits any photometric aperture to $\leq7\farcs1$, however, this aperture appears to encompass much of the ejecta.  We removed the star with three separate attempts using PSF subtraction techniques, one using the nominal PSF provided by the ZTF pipeline, the others using PSFs estimated with independent code.  We masked out strong residuals in the core of the star (6\arcsec{} radius) and measured the brightness of the ejecta in apertures up to 15\farcs2 in radius.  Beyond 11\farcs1, the total brightness was constant or brightened by 0.01~mag per arcsec.  In Table~\ref{tab:outbursts}, we give the average brightness based on the three attempts, which is consistent with all three measurements within 1$\sigma$.

          Our first outburst image appears to taken $\sim$12~hr before the outburst peak as observed by \citet{farnham19-wirtanen-tess}.  Therefore, our peak brightness estimate may be low by 0.1~mag.

    \item[(C)] Scaling and subtracting the 2018 December 10 $g$-band image from the December 12 UTC $g$-band image resulted in a halo of negative residuals around the outburst ejecta, perhaps because our photometric scaling is designed for small apertures yet the extended coma at this time is more affected by gas (i.e., \ch{C2}).  We instead examine the $r$-band data from December 04 and 12.  Based on these images, the outburst appears to have three components at position angles 36, 72, and 296\degr.  The interpretation of the morphology is affected by the subtraction, which leaves strong negative residuals towards PA$\sim$180\degr, and more subtle residuals towards 55\degr{}.  We enhanced the $r$-band images by normalizing them with an azimuthally averaged coma (Fig.~\ref{fig:c-ztf}).  This confirms that the two components at 36 and 72\degr{} are not an artifact caused by over-subtraction of the coma along PA$\sim$55\degr.

          The residual emission is distributed as far as 400\arcsec{} (23,700~km) from the comet.  Aside from an ion tail, it is difficult to ascertain how much of this emission beyond 400\arcsec{} is from the outburst or from residual background.  Therefore we only report photometry within this radius.

          We also inspected the Lowell 0.8-m data taken on 2018 December 05 and on December 12 at 02:07 and 08:46 UTC.  Examination of these data reveals ejecta motion over this 6.65-hr period (Fig.~\ref{fig:c-lowell}).

    \item[(D)] One ZTF $g$-band image taken 2019 January 24 was subtracted from the $g$-band image taken on January 28 UTC.  A small extended source remains in the difference.  It has a v-shaped morphology, reminiscent of event C.  There is a near linear feature, 27\arcsec{} long and pointing towards position angle 188\degr, and a shorter, 21\arcsec{} long, but broader feature pointing towards 240\degr.  Faint arcminute-scale extended emission is present in the residual image, possibly from \ch{C2} gas.

    \item[(E)] After scaling and subtracting three images (1 $g$, 2 $r$) taken on 2019 February 20 from the $r$-band image on 2019 February 24 UTC, a clear residual is detected, no larger than 5\farcs1 in radius.  However, there is possible extended ejecta towards position angles 180 to 270\degr{} in the smoothed contours of the residuals, out to $\sim30$\arcsec.

    \item[(F)] After scaling and subtracting 8 and 11 ZTF images from the data taken on 2019 April 14 and 18 UTC, respectively, we are unable to identify any source in the residuals.

    \item[(G)] Three baseline images, 2 $r$ and 1 $g$ taken 2019 June 15 to 19 UTC, were scaled and subtracted from the first outburst image on 2019 July 4 UTC taken in the $g$-band.  The image of the ejecta is noisy, but residuals are detected out to 18\arcsec.
  \end{itemize}

  \begin{deluxetable*}{lcDDcccDccDc}
    \tablecaption{Summary of outburst circumstances and properties.}
    \label{tab:outbursts}
    \colnumbers
    \tablehead{
      \colhead{Label}
      & \colhead{Date}
      & \multicolumn2c{$(T-T_P)_0$}
      & \multicolumn2c{$(T-T_P)_1$}
      & \colhead{$\Delta t$}
      & \colhead{$\log_{10}H$}
      & \colhead{Filter}
      & \multicolumn2c{$\Delta m_5$}
      & \colhead{$\rho$}
      & \colhead{$m_e$}
      & \multicolumn2c{$G$}
      & \colhead{$M$}
      \\
      & \colhead{(UTC)}
      & \multicolumn2c{(days)}
      & \multicolumn2c{(days)}
      & \colhead{(days)}
      & \colhead{(J \inv[2]{m})}
      &
      & \multicolumn2c{(mag)}
      & \colhead{(\arcsec)}
      & \colhead{(mag)}
      & \multicolumn2c{(km$^2$)}
      & \colhead{(kg)}
    }
    \decimals
    \startdata
    A                  & 2018-08-09 & -131.456 & -125.683 &  \nodata  & \nodata & $r$ & $-0.93$ &   7 & $17.49\pm0.04$ &  26.9 & 3.6\timesten{5} \\
    B\tablenotemark{a} & 2018-09-26 &  -77.83  &  -77.81  &  $51\pm3$ & 9.34    & $i$ & $-1.37$  &  11 & $14.11\pm0.05$\tablenotemark{b} &  77.8 & 1.0\timesten{6} \\
    C                  & 2018-12-12 &   -2.682 &   -0.700 &  $76\pm1$ & 9.80    & $r$ & $-0.51$ & 475 &  $9.06\pm0.02$ & 118. & 1.6\timesten{6} \\
    D                  & 2019-01-28 &   44.436 &   46.234 &  $47\pm2$ & 9.66    & $g$ & $-0.15$ &  32 & $14.26\pm0.03$ &  16.5 & 2.2\timesten{5} \\
    E                  & 2019-02-24 &   69.403 &   73.297 &  $26\pm3$ & 9.26    & $r$ & $-0.21$ &   5 & $17.80\pm0.07$ &   2.5 & 3.3\timesten{4} \\
    G                  & 2019-07-04 &  189.292 &  203.251 & $124\pm9$ & 9.63    & $g$ & $-1.63$ &  18 & $17.04\pm0.13$ & 387. & 5.2\timesten{6} \\
    \enddata
    \tablecomments{Columns: (1) Event label from Fig.~\ref{fig:lightcurve}; (2) Date of first detection; (3) Time of event with respect to perihelion, lower-limit; (4) Time upper-limit; (5) Time since last event and full-range uncertainty; (6) Solar radiant exposure since last event; (7) Observed peak change in brightness as $r$-band magnitude in 5\arcsec{} radius aperture; (8) Filter; (9) Photometric aperture radius; (10) Total brightness of ejecta in the $r$-band and $1\sigma$ absolute uncertainty; (11) Total geometric cross section; (12) Total mass, assuming $dn/da\propto a^{-3.5}$ (see Section~\ref{sec:mass} for details).}

    \tablenotemark{a}{Outburst timing from \citet{farnham19-wirtanen-tess}.}

    \tablenotemark{b}{The lightcurve of \citet{farnham19-wirtanen-tess} suggests the peak brightness is $-0.1$~mag brighter.}
  \end{deluxetable*}

  Two of the outbursts have color measurements on the night of the outburst discovery: 0.47$\pm$0.04~mag for C and 0.50$\pm$0.05~mag for E.  The $g-r$ colors of these events are within 1$\sigma{}$ of the mean coma color within 5\arcsec{} radius photometric apertures.

  \subsection{Search for fragments}
  We used the \hst{} images of comet Wirtanen obtained on December 13 to look for evidence of any fragments that might have been ejected in the December 11/12 outburst.  The close proximity of the comet (0.08 au) and pixel scale of the \hst{} WFC3 images (0.04 arcsec/pix) allowed us to investigate the region within a projected distance of around 2300~km of the nucleus for any lingering material.  Our observations consist of sequences obtained between 11:32 and 16:18 UTC on December 13 (approximately 35 to 40 hours after the onset of the outburst).  Our search utilized four images obtained with the F689M filter and five images with the F845M filter, each with exposures short enough for the comet to be untrailed.  We used the drizzle-processed (DRZ) images, registered on the comet optocenter and rotated so that North was up and East to the left.

  The biggest complication of the search is the large number of cosmic rays that impact the \hst{} observations, mimicking the types of features that we are looking for.  Thus, we used cosmic ray cleaned data in addition to using the (uncleaned) DRZ images.  Although this improved the situation somewhat, a significant number of cosmic rays still remained.  Ultimately, we investigated both versions, in case the cosmic ray removal was also removing fragments.  We also enhanced the images with two different techniques, applying an azimuthal average and a (Gaussian) unsharp mask that removes the bright central peak of the comet and improves the contrast of any fragments.

  In order to constrain our search, we assumed that any fragments must be moving slowly enough to remain in the field of view for 40 hours (the time from the onset of the outburst to the last \hst{} observation in this set), setting an upper limit on the proper motion of 25~pix~\inv{hr} (a projected velocity of 16~\mps{} at the comet).  We also assumed that particles large enough to be detected will not accelerate significantly during the 5-hr window of the \hst{} images, and thus any candidates will move along a line with spacing proportional to the intervals in the observation times.

  For each combination of filter/enhancement, we blinked the sequence of images to look for candidate particles with acceptable motions.  In another approach, we co-added the sequences from each filter (and processed as needed), allowing us to look for linear strings of particles that would represent a moving fragment.  In all of our searches, we found no convincing evidence for fragments in the \hst{} images.

  Using the cosmic rays as a guide, we estimate that we should have detected any point source or central condensation that produces a signal of at least 2\timesten{-18}~W~\inv[2]{m}~\inv{\micron} (F689M, 0.5~electrons~\inv{s}).  If we assume an inactive spherical shape with 4\% albedo, then our detection limit suggests that we should see any fragment larger than $\sim$2~m in radius, or a mini-comet with a dust cross sectional area of $\sim$12~m$^2$. (These estimates ignore issues such as phase effects, but these are small relative to other uncertainties.)

  \section{Analysis}\label{sec:analysis}
  \subsection{Ejecta expansion, grain size}
  Our general assumption is that all outbursts are brief events, lasting $\ll1$~day, and that the ejecta can continue to be observed well after the outburst is over.  This assumption is consistent with the analysis of 30-min cadence observations of outburst B with the \textit{Transiting Exoplanet Survey Spacecraft} (\textit{TESS}) by \citet{farnham19-wirtanen-tess}, who found coma brightening ceased after 8~hr.  Short outburst timescales, $\lesssim 1$~hr, are also consistent with the high spatial resolution observations of outbursts at 9P/Tempel~1 by \di{} \citep{farnham07} and at 67P/Churyumov-Gerasimenko by \rosetta{} \citep{knollenberg16-outburst, vincent16-fireworks, agarwal17-outburst, bockelee-morvan17-outbursts, rinaldi18-outbursts}.  Therefore, most outbursts sampled with a 1 to 3~day cadence and a small aperture will have an observation near or after the peak.

  The fact that event E does not have a distinct photometric peak suggests the peak occurred within the 3.9-day gap in data, and that ejecta has moved outside our nominal (5\arcsec{}) photometric aperture.  In Section~\ref{sec:results}, we identified faint extended emission in the outburst residuals, up to $\sim$30\arcsec{} from the nucleus, consistent with this possibility.  To illustrate, an expansion speed of 50~\mps{} and a projected distance of 30\arcsec{} corresponds to an outburst time 2.6~days before the first observation of event E, comfortably within the 3.9-day gap in photometry.

\begin{figure*}
    \plottwo{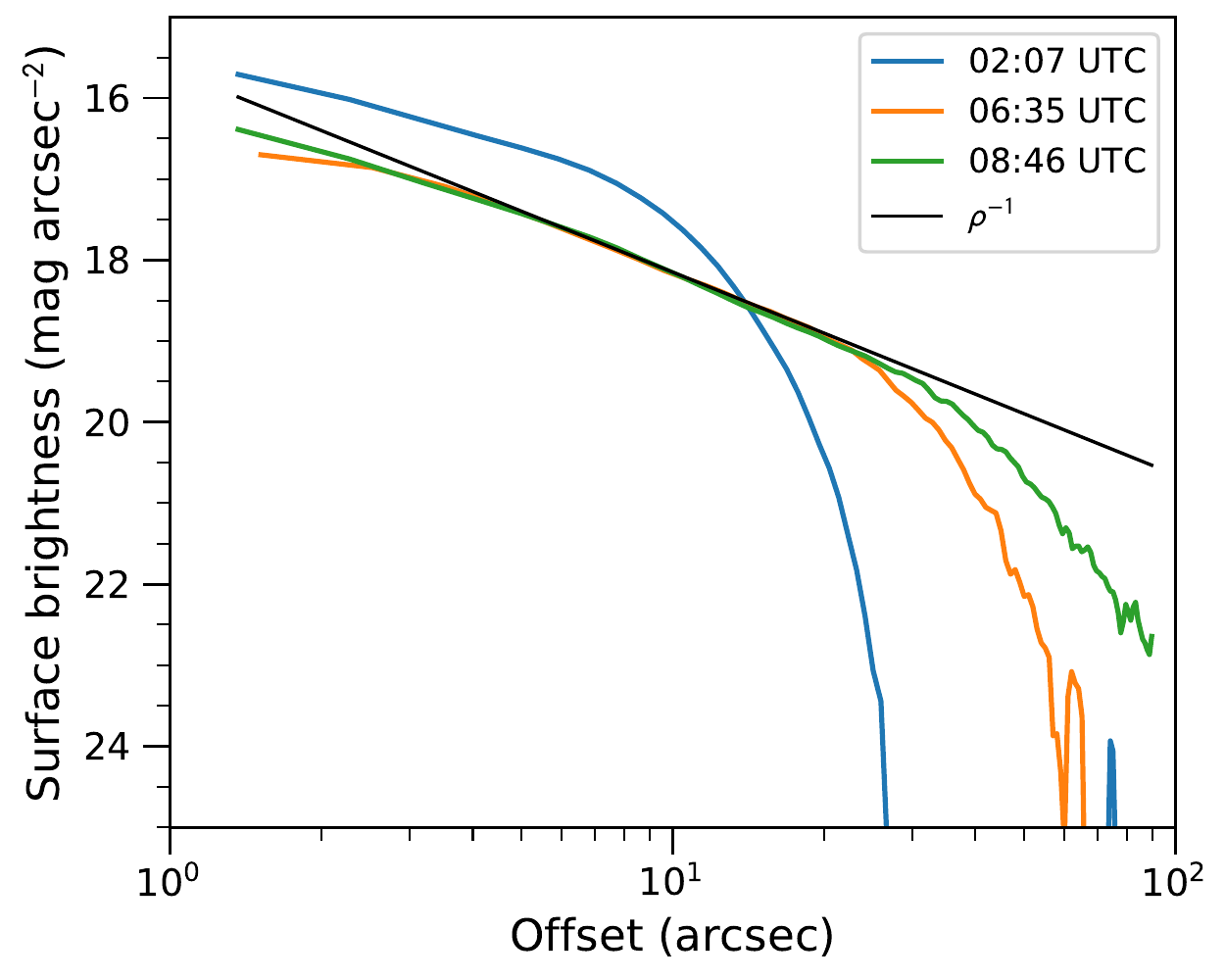}{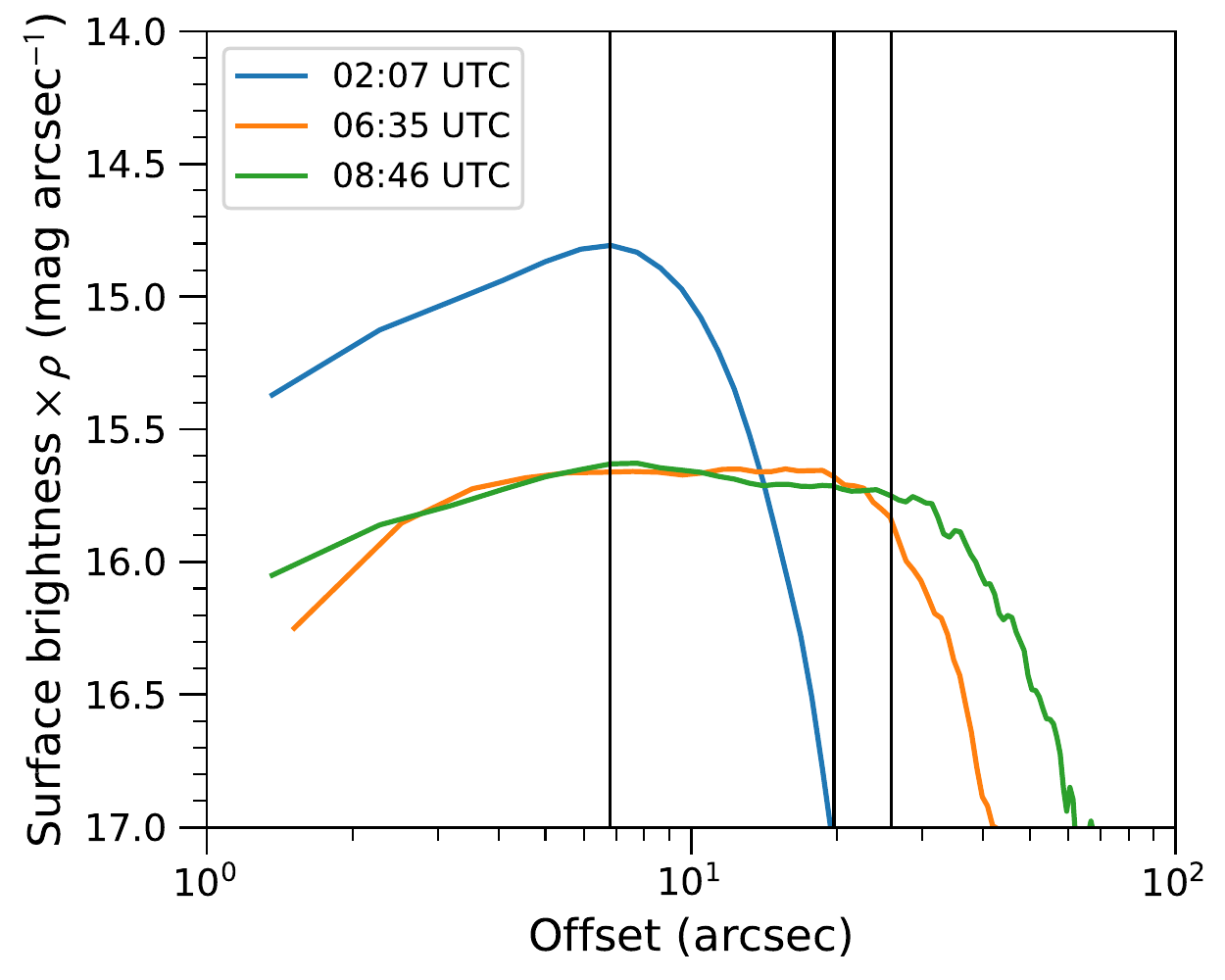}
    \caption{(Left) Outburst C ejecta surface brightness based on Lowell 0.8-m and ZTF images.  The images are sampled with a 5-pix wide line along the anti-sunward direction, i.e., along the horizontal feature in Fig.~\ref{fig:c-lowell}. (Right) Surface brightness profiles multiplied by distance to the nucleus ($\rho$).  Vertical lines mark our estimates of the leading edge.}
    \label{fig:c-expansion}
  \end{figure*}

  For outburst C, we showed motion in the ejecta over a 6.65-hr period.  Of the two features identified in Fig.~\ref{fig:c-lowell}, the anti-sunward feature is brighter and easier to measure.  In Fig.~\ref{fig:c-expansion}, we plot the surface brightness of the ejecta measured in a 5-pixel-wide box along the anti-sunward direction in the Lowell and ZTF $r$-band data.  Each profile is nearly linear in log-log space closest to the nucleus, then falls with respect to this line at farther distances.  We use the break point (manually estimated in profiles multiplied by $\rho$, Fig.~\ref{fig:c-expansion}, right) to measure the motion of the material.  For break points at 6\farcs8, 19\farcs7, and 29\farcs5, and assuming 1-pix uncertainties, the expansion speed based on a linear fit is 55.1$\pm$3.1~\mps, and an outburst age of 21.3$\pm$0.9~hr in our first image.  This places an approximate outburst onset at December 11 04:49 UTC ($^{+1.9}_{-1.6}$~hr).  The reduced-$\chi^2$ statistic is 3.8, but with only 1 degree of freedom there is a 5\% probability of having reduced-$\chi^2\geq3.8$ \citep{bevington92}.  Thus, we conclude a non-linear expansion is possible, but not strongly supported by our data.

  An upper limit on the outburst C ejecta speed can be estimated from the extent of the residuals in the ZTF image (400\arcsec) and the estimated start time of the outburst.  Together, they yield an expansion speed of 250~\mps.

  The lack of outburst ejecta in the \hst{} images suggests a lower limit to the expansion speed, assuming any slowly moving material is not too diffuse to identify.  Given the 26-hr gap between the last Lowell 0.8-m image and the first \hst{} image, and that the comet is about 40\arcsec{} from the image edge in the anti-sunward direction, the slowest ejecta moved faster than $\sim$23~\mps{} in projection on the sky.

  For the anti-sunward material in outburst C, we can consider the effects of radiation pressure and estimate a lower limit on the grain size assuming the material is in linear expansion.  \citet{burns79} present the acceleration due to solar radiation pressure, $a_{\mathrm{rad}}$ on a dust grain as
  \begin{equation}
    a_{\mathrm{rad}} = \frac{Q_{\mathrm{pr}}L_\sun G}{4 \pi \rh^2 c m},
  \end{equation}
  where $Q_{\mathrm{pr}}$ is the grain radiation pressure efficiency averaged over the solar spectrum, $L_\sun$ is the Sun's total luminosity \citep[nominal value 3.828\timesten{26}~W;][]{prsa16-constants}, $G$ is the grain geometric cross sectional area, $c$ is the speed of light, and $m$ is the mass of the grain.  For simplicity, we take $Q_{\mathrm{pr}}=1$.  The projected acceleration on the sky is $a_{\mathrm{rad}}$ attenuated by $\sin(\theta)$, where $\theta$ is the Sun-target-observer angle.  Grains are accelerated (5.3\inv{$a$})~\mps~\inv{hr} in the anti-sunward direction, projected onto the plane of the sky, where $a$ is the grain radius in \micron.  This acceleration corresponds to a total displacement of (144\inv{$a$})~km between the ZTF and second Lowell 0.8-m epoch (06:35 and 08:46 UT), or at about the level of the seeing (59~km~\inv{arcsec}) for 1~\micron{} grains.  Therefore, the optically dominant grains in this feature are likely at least 1~\micron{} in radius.

  \subsection{Total geometric cross-sectional area and outburst mass}\label{sec:mass}
  Converting the observational data into physical quantities allows us to make meaningful comparisons between each outburst and the ambient coma.  However, this conversion relies upon several unknown quantities, and therefore will be dependent on our adopted parameters and assumptions.  First, we assume a dust $V$-band geometric albedo of $A_p(V)=4.00\%$.  Given our measured colors, the corresponding albedos are 3.82, 4.19, and 4.22\% at $g$, $r$, and $i$, respectively.  Ignoring the dependence of scattering efficiency on grain size, the total geometric cross-sectional area, $G$, within a photometric aperture is
  \begin{equation}
    G = \frac{\pi \rh^2 \Delta^2}{A_p \Phi(\theta)} 10^{-0.4 (m - m_\sun)},
  \end{equation}
  where $\Delta$ is the observer-comet distance in units of length, $\Phi(\theta)$ is the coma phase function evaluated at phase angle $\theta$, $m$ is the apparent magnitude of the dust, and $m_\sun$ is the apparent magnitude of the Sun at 1~au in the same bandpass and magnitude system.  For \rh{} expressed in units of au, $G$ will carry the units of $\Delta^2$.  The coma and outburst photometry are converted to $G$ and listed in Tables~\ref{tab:obs} and \ref{tab:outbursts}, respectively.

  Converting cross-sectional area to dust mass is more uncertain.  Here, we require assumptions on the grain density and grain size distribution.  For density, we take 1000~kg~\inv[3]{m}, which allows for some porosity in the grains.  Power-law size distributions roughly approximate the grain size distributions observed in situ by spacecraft dust instruments and impacts on the \stardust{} collector \citep{mcdonnell87, green04, price10-stardust, fulle16-giada, merouane17-cosima}.  We assume a differential size distribution, $dn/da$, with a power-law slope of $k=-3.5$, which is within the estimated time-averaged value of $-3.3\pm0.3$ derived by \citet{fulle00-wirtanen} from 46P's coma morphology.  It is also the cross-over point for mass estimates based on observed brightness, i.e., for values $>-3.0$ the largest particles dominate the estimated mass, whereas for $<-4.0$ the smallest particles dominate the mass.  Finally, we assume the dust grain radii span from 0.1~\micron{} to 1~mm.  For these parameters, we convert the outburst geometric cross-sectional area estimates to total mass and provide them in Table~\ref{tab:outbursts}.  The masses range from 3\timesten4 to 5\timesten6~kg.  For $k=-3$, increase the mass estimate by a factor of 10, for $k=-4$, decrease the estimate by a factor of 10 \citep[e.g., see][]{tubiana15}.

  \subsection{Lack of Boulders in Outburst Ejecta}

  In the \hst{} images, there was no evidence for ejecta from outburst C, including point sources.  The lack of boulder-sized ejecta may be because: none were ejected, they were smaller than $\sim$2~m in radius, they moved faster than 23~\mps{} in the plane of the sky, or they disintegrated before \hubble{} could observe them.  Whether or not any fragments larger than 2~m were ejected is difficult to assess.  The mass of a 2-m radius chunk of nucleus would be 2\timesten{4}~kg, assuming a density of 500~kg~\inv[3]{m} for the nucleus \citep[similar to comet 67P;][]{jorda16}, well within the mass budget of the outburst (nominally 2\timesten{6}~kg, Table~\ref{tab:outbursts}).  However, fragments may be no larger than $\sim10$~m, which have a mass of 2\timesten{6}~kg.  Note, these arguments assume a constant power-law from small grains to macroscopic fragments.

  Fragments with sizes near 10~m in radius have been observed in cometary comae, with substantially long lifetimes.  An outburst of fragment B of comet 73P/Schwassmann-Wachmann~3 in 2006 produced mini-comets up to $R\sim10-100$~m with lifetimes of at least a month \citep{fuse07, ishiguro09-sw3}, and small ($R\lesssim30$~m) fragments of comet 332P/Ikeya-Murakami survived for at least a few months \citep{jewitt16-332p}.  We can estimate the lifetime of meter-size fragments, by considering the effects of sublimation: erosion and rotational spin up to fragmentation.

  The sublimation rate of water ice at 1.0~au in contact with low-albedo material (Bond albedo of 5\%, i.e., a cometary surface) is $3.6\timesten{17}$~\mpscc{} for a slowly rotating sphere \citep[estimated following][]{cowan79}. Assuming an ice-to-dust mass ratio of 0.2 \citep[e.g.,][]{rotundi15}, and a 1:1 mixture of silicates (3300~kg~\inv[3]{m}) and carbonaceous dust (1500~kg~\inv[3]{m}) \citep{bardyn17-carbon, woodward21-c2013us10}, the mean erosion rate is 9~cm~\inv{day}.  This estimate assumes a 100\% active fraction (water production rate / water ice sublimation rate), whereas comet active fractions are typically 10\% or less \citep{ahearn95}, reducing the erosion rate to 9~mm~\inv{day}.  Thus the lifetime of a meter-sized fragment due to water ice driven erosion may be about 100 days, but not much less than 10 days.

  Rotional spin up of mini-fragments to disintegration has been previously considered.  \citet{jewitt20-borisov} estimate a few hours to a day for a 1~m object at \rh=1~au, based on the (scaled) torque imparted on the nucleus of 9P/Tempel~1, as estimated by \citet{belton11}.  \citet{steckloff16-syorp} use the YORP formalism to estimate the sublimation driven spin up to disintegration in order to describe the formation of tail striae.  Based on their approach, we compute a timescale of at least 75~days for a 2-m object at 1~au.  While it is possible that large fragments could have disintegrated in the 35 to 40 hours after ejection from the nucleus, we could not identify any ejecta material in the \hst{} images at all, whether produced by the outburst itself, or by the subsequent fragmentation of cometary boulders.  Our preferred conclusion is that no large ($>$2 m) boulders were ejected.

  \subsection{Outburst Frequency}\label{sec:frequency}
\begin{figure}
    \plotone{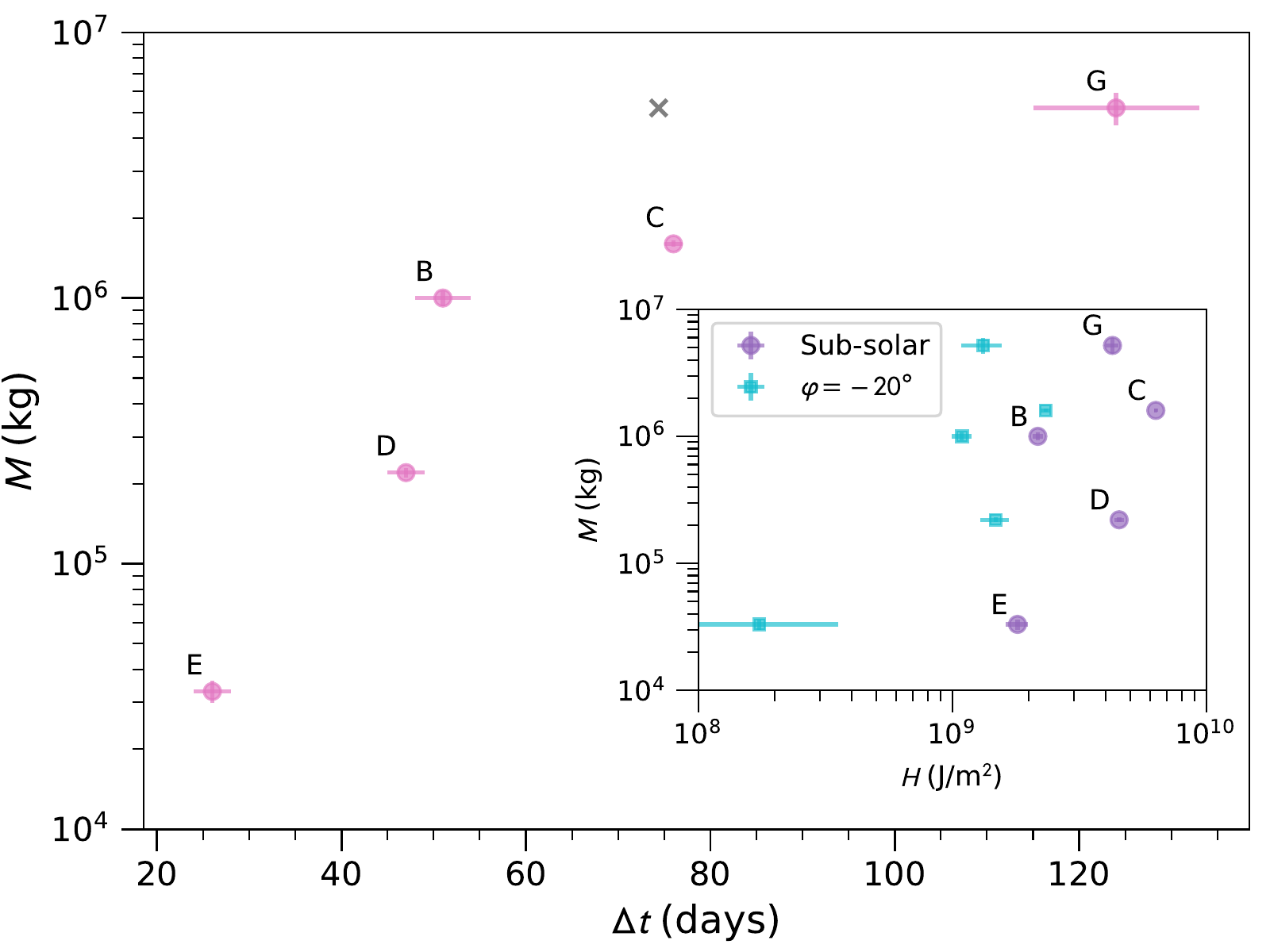}
    \caption{Estimated outburst mass (nominal grain parameters) versus time elapsed since last outburst.  The $\times$ symbol marks the location of outburst G if lightcurve anomaly F (Fig.~\ref{fig:lightcurve}) was considered to be an outburst.   (Inset) Mass versus solar radiant exposure since last outburst.  The exposure is calculated for the sub-solar point and for a source at --20\degr{} latitude.}
    \label{fig:mass-correlation}
  \end{figure}

  The six outbursts occur throughout the observed period.  Neglecting the significant gaps in the lightcurve where small events may have taken place (especially near $-40$, +80, and +160~days), we list the time elapsed between each outburst, $\Delta t$, in Table~\ref{tab:outbursts} and plot ejecta mass versus $\Delta t$ in Fig.~\ref{fig:mass-correlation}.  There is an intriguing correlation between $\Delta t$ and the amount of material ejected.  Pearson's correlation coefficient calculated for $\Delta t$ and $\log_{10}M$ is 0.89, indicating a strong significance.

  In an attempt to better understand the cause of the apparent correlation, we estimated the solar radiant exposure, $H$, based on the comet-Sun distance over the time periods between outbursts, and list them in Table~\ref{tab:outbursts} \citep[assuming a solar luminosity of 3.828\timesten{26}~W;][]{prsa16-constants}.  The correlation is not as good (0.58), as seen in Fig.~\ref{fig:mass-correlation}.  All five outbursts occur within a narrow range of radiant exposures, from 2 to 6\timesten{9}~J~\inv[2]{m}, despite spanning two orders of magnitude in mass / cross-sectional area.  However, our radiant exposure calculation does not consider the source location and pole orientation, nor local topography.

  With a pole solution and the assumption of a spherical nucleus, we can explore if a single source illuminated by the Sun could be responsible for all six outbursts.  The best pole solutions of \citet{knight21-wirtanen} indicate a high obliquity, with equinox near perihelion.  Thus, a near equatorial source could be illuminated during each outburst.  For their best pole solution, RA, Dec = 319\degr, --5\degr{} (obliquity of 70\degr), we find that planetocentric latitudes from --20\degr{} to +30\degr{} are illuminated during outbursts A through G (5\degr{} steps were tested).

  We re-calculated the solar radiant energy, this time considering a single source region on a rotating spherical nucleus with the pole orientation of 46P from \citet{knight21-wirtanen} and latitudes from --20\degr{} to +30\degr{}.  We searched for solutions that would improve the mass-energy correlation.  More southern latitudes greatly reduced the amount of energy received before outburst E occurred.  We show a latitude of --20\degr{} in Fig.~\ref{fig:mass-correlation} as an example.  Due to the change in energy for event E, the correlation coefficient between $\log_{10}M$ and $\log_{10}H$ increased from 0.58 to 0.79.  However, the scatter between events B, C, D, and G are not improved.  This exercise does not demonstrate that these events are all physically connected, but assuming that they are, insolation is likely not responsible for the correlation between ejecta mass and time since the last event.

  \section{Discussion}\label{sec:discussion}

  \subsection{Other Observations of Wirtanen's Outbursts}
  \citet{farnham19-wirtanen-tess} and \citep{farnham21-wirtanen} observed outbursts B, C, and D.  \citet{farnham19-wirtanen-tess} analyzed \textit{TESS} observations of outburst B.  They found dust expansion speeds of a few tens of m~\inv{s}, based on the size of their photometric aperture (25,000~km radius) and the centrally condensed appearance in the images (7900~km~\inv{pix}) that lasted up to 20 days.  \citet{farnham21-wirtanen} observed outbursts C and D in ground-based images with a near 1-hr cadence, allowing them to estimate dust expansion speeds of 68$\pm$5 and 162$\pm$15~\mps, respectively.  Our temporal resolution is coarser than that of \citet{farnham21-wirtanen}, but our estimated expansion speed for the anti-sunward ejecta in outburst C, 55$\pm$3~\mps{}, is in agreement.  The fast moving material in outburst C (250~\mps) is less than a factor of two faster than the \citet{farnham21-wirtanen} measurement of outburst D.

  \citet{combi20-soho} and \citet{combi20-soho-pds} analyzed SOHO/SWAN observations of comet Wirtanen's atomic hydrogen coma in order to estimate water production rates, and identified two post-perihelion outbursts in the 2002 apparition, with peaks at 15 and 36~days after perihelion.  The effective water production rates increased by a factor 4 to 5, but note that the photometric aperture is large (8\degr{} radius) and includes $\sim$2--3 days of activity.  Assuming 2 days of averaging, the effective number of water molecules from each of the outbursts is about $\sim$1\timesten{34}, or about 3\timesten{8}~kg.  These are significantly larger then what we observed in 2018 by two to three orders of magnitude.  The frequency of these large events over the observed 75 day period is 0.027~day$^{-1}$, compared to our rate of 6 in 352 days (0.017~day$^{-1}$).  However, if we consider the entire SOHO data set, which has good temporal coverage for 209 days spread out over four perihelion passages (1997, 2002, 2008, and 2018), the rate becomes 0.0096~day$^{-1}$, or about one large event per perihelion passage.

  Unfortunately, none of our outbursts are covered by their 2018/2019 data set.  There are three anomalously high ($\sim2\sigma$) points in the 2018 water production rate time series near 29.15 to 31.15 days after perihelion.  Our single photometry point (a 2.3$\sigma$ outlier) at 30.32 days does not confirm any dust outburst at that time.

  The large outburst frequency based on the SOHO data ($\sim$1 per perihelion passage) is borne out in optical lightcurves of comet Wirtanen: \citet{kidger08} reports a $-$2-mag outburst in a 10\arcsec{} radius aperture (1.2\timesten{4}~km) 103~days after perihelion, with good temporal coverage over 220 days; in the assembled lightcurve by \citet{yoshida13-46p-2002}, a $\sim-4$~mag outburst\footnote{First observed by K.\ Kadota (Ageo, Japan): \url{https://groups.io/g/comets-ml/message/2585}.} is apparent 29~days after the 2002 perihelion (this is likely the same as the second event observed by SOHO); \citet{kidger04} have sparsely sampled data in 2002, but suggest another possible outburst near 215~days after perihelion (observed after a $\sim$70-day gap in coverage).  This may be coincident with our outburst G, except it is separated by two orbits.

  \subsection{Mini-outbursts of Wirtanen and Other Comets}
  In terms of mass, the outbursts of comet Wirtanen are similar to the mini-outbursts of comet 9P/Tempel~1 and 67P/Churyumov-Gerasimenko.  At comet 67P, the ejecta mass estimates are of order $10^4$ to $10^5$~kg, based on the analysis of 34 outbursts by \citet{vincent16-fireworks}.  They also re-analyzed the 2005 July 02 outburst of comet 9P, and, with the same assumptions and techniques, estimated a mass of 5\timesten5~kg.  Other mini-outbursts of 9P are the same order of magnitude or smaller \citep{farnham07}.  With the grain parameters of \citet{vincent16-fireworks}, $dn/da=a^{-2.6}$ for 1--50~\micron{} in radius, we recalculated the ejecta masses of the Wirtanen outbursts: 5.9\timesten{4} to 9.3\timesten{6}~kg (events E and G, respectively).  Thus the Wirtanen outbursts are the same order of magnitude to one order larger than the events at 9P and 67P.

  \citet{vincent16-fireworks} estimated the source locations for the 67P mini-outbursts, and found they were correlated with regional boundaries, especially near steep scarps or cliffs.  Indeed, \citet{grun16} correlated an outburst to sunrise on a cliff, \citet{pajola17} directly connected an outburst to an observed cliff collapse, and \citet{agarwal17-outburst} associated an outburst with the collapse of an overhanging wall.

  At comet 9P, a correlation with areas of high topographical relief or pits has been suggested by \citet{belton08}.  The relationship is intriguing but uncertain.  \citet{belton08} analyzed broad ejecta patterns back to planetocentric coordinates of an unresolved nucleus, whereas \citet{vincent16-fireworks} worked with nucleus-resolved data, and in some circumstances could visually pinpoint the outburst source to the pixel level.  The techniques of \citet{belton08} inherently assume the nucleus is spherical and the outburst ejected normal to the surface.  However, 9P is faceted, and many of these facets face the same direction.  Therefore, the projection of planetocentric coordinates to the shape model is multi-valued, and the source regions for the 9P mini-outbursts are uncertain.

  No outburst equivalent to those seen at 67P, 9P, and 46P was observed at 103P/Hartley 2.  \citet{meech11-epoxi} note an outburst of 103P on 2010 September 16 based on water production rates but without additional details and the event was not seen in \textit{SOHO} observations of the H$\alpha$ coma \citep{combi11-h2}.  \citet{lin13} tentatively associate a relative change in jet brightness in processed data with an outburst, but also consider changes in grain properties as a possibility.  We note that the comet's lightcurve, as observed by \textit{Deep Impact}, has three-peak pattern during this period \citep{ahearn11, bodewits18-h2}, and that the time of the change observed by \citet{lin13} corresponds to the brightest of the three peaks.

  We take the analyses of the 67P mini-outbursts as a guide, and assume most or all mini-outbursts are related to steep scarps, cliffs, and other features of high topography.  If true, then the differences in outburst frequency between 67P, 9P, and 103P are related to differences in terrain.  That is, the paucity of large cliffs, etc.\ on the nucleus of 103P results in a lack of mini-outbursts by that comet.

  We compare the surface area normalized observed frequency of mini-outbursts at comets 67P, 9P, 46P, and 103P in Table~\ref{tab:frequency} (references to outburst rates and nuclear surface area are contained therein).  An upper-limit to the outburst frequency of comet 103P is based on the lack of outbursts observed for this comet during the 2010 perihelion.  No outbursts were observed over the 180-day lightcurve of \citet{meech11-epoxi}.  If 103P had the same rate of mini-outbursts as 46P, 2 to 3 events could have been seen, but details on whether or not they would have been detected depend on observing circumstances and cadence.  Perhaps the most sensitive monitoring was executed with the \textit{Deep Impact} spacecraft over a period of approximately 3 months \citep{ahearn11}, but without any reported events (1 to 2 expected).  Thus, we estimate 103P's mini-outburst rate to be no more than that of 46P, or $\lesssim$0.004~\inv{day}~\inv[2]{km}.  We find that comet 67P and 9P have outburst frequencies two orders of mangitude larger than those of 46P and 103P.  Most of this difference is due to the large nuclear surface areas of 67P and 9P.  However, the area normalized rates are still $\sim$3 to 5 times larger than those of 46P and 103P.  If 46P had the same area normalized outburst frequency as 67P, then we could have seen 27 outbursts in our dataset.

  \begin{deluxetable}{lcccb{5cm}}
    \tablecaption{Summary of mini-outburst frequencies.}
    \label{tab:frequency}
    \tablehead{
      \colhead{Comet}
      & \colhead{$f$}
      & \colhead{$A$}
      & \colhead{$f/A$}
      & \colhead{References}
      \\
      & \colhead{(\inv{day})}
      & \colhead{(km$^2$)}
      & \colhead{(\inv{day}~\inv[2]{km})}
    }
    \startdata
    67P/Churyumov-Gerasimenko & 0.8 & 46.9 & 0.02 & \citealt{vincent16-fireworks, jorda16} \\\hline
    9P/Tempel 1 & 1.2 & 108 & 0.011 & \citealt{belton08, thomas13-tempel1}  \\\hline
    46P/Wirtanen & 0.017 & 3.94 & 0.0043 & \citealt{boehnhardt02-wirtanen}; this work  \\\hline
    103P/Hartley 2 & $<$0.02 & 5.24 & $<$0.004 & \citealt{ahearn11, meech11-epoxi, thomas13-hartley2}; this work \\
    \enddata
    \tablecomments{$f$, outburst frequency; $A$ nuclear surface area.}
  \end{deluxetable}

  \citet{vincent17-evolution} identified a correlation between nuclear surface topography and insolation at comet 67P.  Based on an analysis of cliff heights ($\sim$10--100-m scale), they found that regions exposed to more sunlight have fewer large cliffs, and proposed that the erosion of surfaces relaxes their topographies.  They continued by analyzing the surfaces of other comets visited by spacecraft, and suggested an evolutionary sequence from comets 81P/Wild~2 and 67P (roughest), to 9P (intermediate), and finally to 103P (smoothest).  \citet{kokotanekova18-nuclei} hypothesized a similar sequence, based on a correlation between nuclear phase function and albedo.  We build upon these results, adding the correlation between outbursts and cliffs and steep scarps at 67P, and propose that the frequency of mini-outbursts is also correlated with topography.  With respect to surface topography and erosion, comet 46P appears to be in a evolutionary state intermediate to 103P and 9P.

  The interpretation that 46P is similar to 103P in terms of surface topography and evolution relies on the assumption that comet 9P's outbursts are related to steep topography, and that 103P's lack of outbursts is due to it's smoother terrain.  In other words, we assume that the mini-outbursts of 67P are representative of mini-outbursts on all comets.  Potentially 46P's outbursts are instead caused by another mechanism (Section~\ref{sec:intro}).  In Section~\ref{sec:frequency}, we attempted to understand if the outbursts could be caused by a single physically connected system driven by insolation, but we could not account for the apparent correlation between ejecta mass and time delay.  However, we can only approximate the illumination conditions of the surface with a spherical nucleus, so our energy calculations may not be relevant.  Furthermore, there may be non-linear effects, where a small amount of input energy releases a substantial amount of stored energy.  An example of this latter point may be found in the study of a mini-outburst observed at 67P by \citet{agarwal17-outburst}.  They found that a cliff collapse or crack formation likely initiated the event, but that the dust mass loading and speed was inconsistent with free sublimation of ices.  They posited that a sub-surface pressurized gas bubble or the exothermic amorphous-to-crystalline phase transition of water ice provided additional energy to the event.  The kinetic energy per unit mass has been used as metric to test the origins of cometary outbursts.  \citet{ishiguro16-outbursts} found that the kinetic energy per unit mass for outbursts of 15P/Finlay, 17P/Holmes, and 332P/Ikeya-Murakami are $\sim10^4$~J~\inv{kg}, suggesting similar causes.  Thus, studying the energetics and/or dust-to-gas ratio of the 46P outbursts may help discern the driving mechanisms.  However, for small outbursts acceleration from the ambient gas may be need to be accounted for \citep[some considerations for ambient coma are given by][]{gicquel17-outburst}.

  What remains to be addressed is the difference between the circumstances of discovery for the mini-outbursts.  Comet 67P's and 9P's mini-outbursts were primarily observed by spacecraft.  However, two events were observed from the Earth: the 2005 June 14 outburst of 9P observed by \citet{lara06} and \citet{feldman07} (see also the summary by \citealt{meech05}), and a tentative outburst on 2015 August 23 at comet 67P identified by \citet{boehnhardt16}.  The lack of events observed at 67P from the Earth, despite the intensive photometric monitoring of that comet \citep{snodgrass17-campaign}, can be explained by observing geometry and quiescent activity levels.  Setting aside the dependence on observation cadence, the discoverability of an outburst, $D$, is inversely proportional to the scattering cross-sectional area of dust in an aperture, i.e., the $Af$ term in \afr{}.  Outbursts are also more readily discovered at high spatial resolution, which reduces the amount of ambient coma in favor of the point-source like outburst ejecta.  Let $\rho$ be inversely proportional to observer-comet distance $\Delta$ (i.e., fixed angular sized apertures), then the discoverability of outbursts is
  \begin{equation}
    D \propto \frac{\rho}{\afr} \propto \frac{1}{\Delta \afr}.
    \label{eq:discoverability}
  \end{equation}
  For the observational parameters of both comets near perihelion, 1.8~au and 1000~cm for 67P \citep{snodgrass17-campaign}, and 0.08~au and 300~cm for 46P (this work), the ratio is $D(\mathrm{46P}) / D(\mathrm{67P})=75$.  A $-1$-mag outburst of 46P at perihelion in 2018 (i.e., outburst C) would correspond to a $-0.03$-mag outburst of 67P in 2015 at its perihelion, assuming the same dust physical parameters and photometric aperture angular radius.

  \subsection{Potential for Future Mini-Outburst Studies}
\begin{figure}
    \plotone{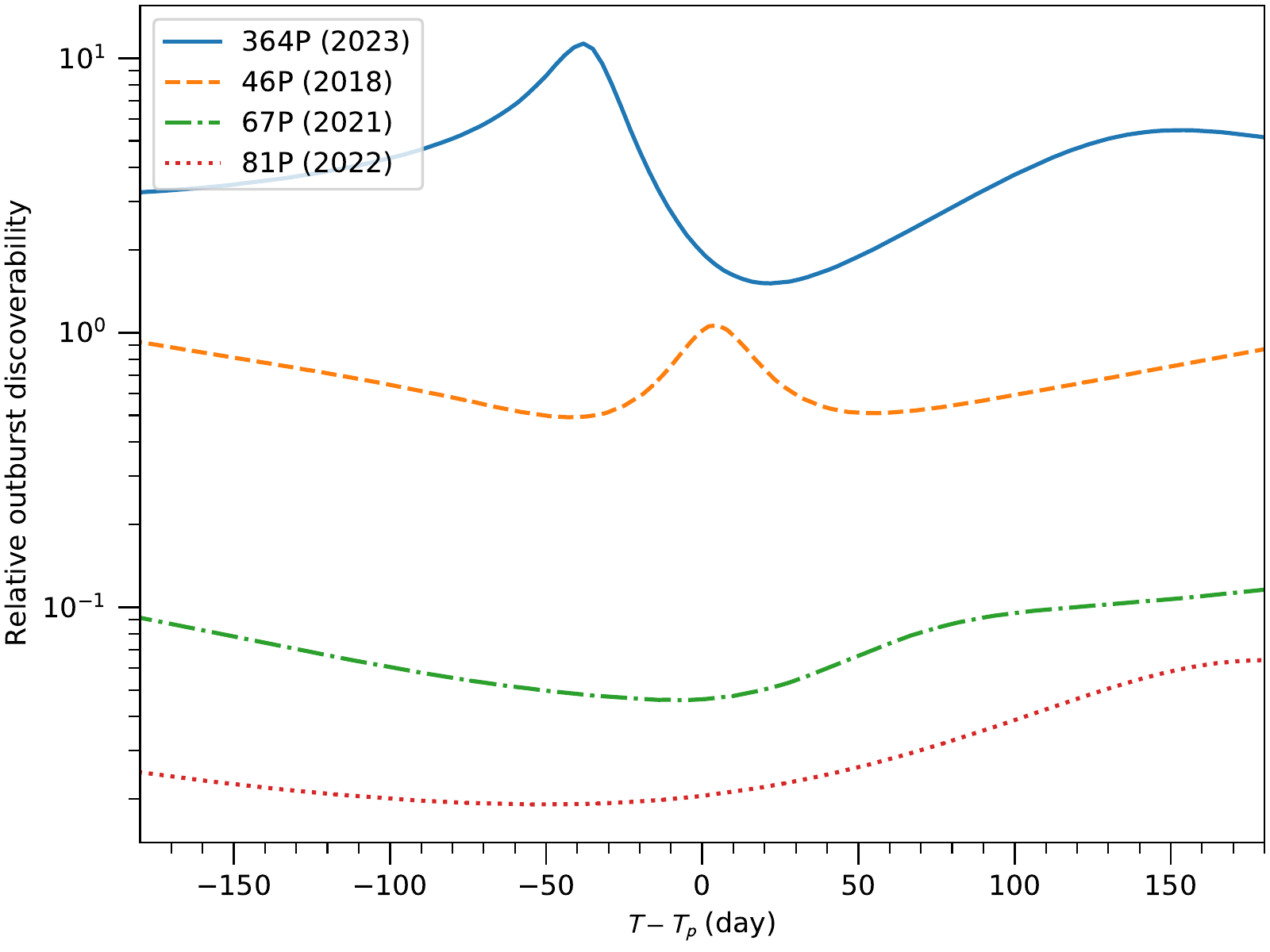}
    \caption{Relative outburst discoverability (Eq.~\ref{eq:discoverability}) versus time from perihelion ($T-T_p$) for four comets.  Discoverability is normalized to that of 46P at perihelion in 2018, thus the same dust cross-section that produces a $-1$~mag outburst at 46P at perihelion would approximately produce a $-0.07$~mag outburst at 67P at perihelion in 2021.  Comets and their perihelion years: 364P/PanSTARRS (2023), 46P/Wirtanen (2018), 67P/Churyumov-Gerasimenko (2021), and 81P/Wild 2 (2022).  Observability of each target, e.g., solar elongation and brightness, has been ignored.}
    \label{fig:discoverability}
  \end{figure}

  The hypothesis that mini-outburst frequency is correlated with surface topography could be tested with comet 81P/Wild~2, which has many cliffs, pits, and rough surface features and a surface area similar to 67P \citep{brownlee04, vincent17-evolution}.  Therefore, this comet may have a mini-outburst every few days.  However, comet 46P/Wirtanen in 2018/2019 provided favorable circumstances for the study of cometary mini-outbursts, and we expect that outbursts of 81P at perihelion in 2022 would be $\sim$50 to $\sim$70 times more difficult to detect \citep[based on the \afr{} measurements of 81P by][]{farnham05}, which may require creative solutions in order to execute such a study.  Close approaches to Earth are great opportunities for mini-outburst discovery, but 81P will be no closer than 0.65~au from the Earth in the next 100 years (JPL Horizons orbit solution K162/9).  The next expected cometary close approach to Earth with a distance similar to 46P will be 364P/PanSTARRS in April 2023 (0.12~au, via the Center for Near-Earth Object Studies\footnote{\url{https://cneos.jpl.nasa.gov/ca/}}), but low solar elongations (minimum 45\degr) will affect the post-approach observability.  To illustrate the differences, we plot the relative discoverability of outbursts at comets 67P in 2021/2022, 81P in 2022/2023, 46P in 2018/2019, and 364P in 2023, for 360~days about perihelion in Fig.~\ref{fig:discoverability}.  \afrho{} values and their variation with heliocentric distance are approximated from results in the literature \citep{farnham05,pozuelos14-wild2-hartley2,snodgrass17-campaign,boehnhardt16}, except for 364P, which we based on the Minor Planet Center photometry database (\afrho[0\degr]=25~cm for $\rho$=10\arcsec, approximates small-aperture photometry reported near perihelion).

  Outbursts are common events with a wide range of strengths \citep{ishiguro16-outbursts}.  Discoveries of outbursts are becoming more prevalent in recent years due to our increased ability to monitor comets (both in the professional and amateur communities), and with the increased efficiencies of survey telescopes and precise all-sky photometric catalogs \citep[e.g., PS1 and SkyMapper;][]{tonry18-refcat2, christian18-skymapper}.  Together, these advances increase our discovery efficiencies, and allow us to identify fainter events.  We expect that current and future cometary outburst surveys will continue to reveal information about cometary behavior and the evolution of cometary surfaces.

  \section{Summary}\label{sec:summary}
  We identified six outbursts in a year-long lightcurve of comet 46P/Wirtanen, with brightnesses ranging from $-0.2$ to $-1.6$~mag with respect to the quiescent trend of the coma, as measured in 5\arcsec{} radius apertures.  The total geometric cross sectional area of dust in the ejecta ranged from 3 to 390~km$^2$, assuming sunlight scattered according to the Schleicher-Marcus phase function.  These areas correspond to $10^4$ to $10^6$~kg of dust, but with a factor of 10 uncertainty due to the unknown grain size distribution.  The mass estimates are similar to or one order of magnitude larger than the mini-outbursts observed at comets 9P/Tempel~1 and 67P/Churyumov-Gerasimenko.

  The expansion speed of material ejected by an outburst near perihelion was at least $55\pm3$~\mps{} and up to 250~\mps{}, projected to the plane of the sky.  \textit{Hubble Space Telescope} images taken $<$2~days after the start of this outburst lack any sign of macroscopic fragments ($\sim$2-m lower limit radius), or any ejecta at all, indicating a minimum ejection speed of 23~\mps{}.

  The time difference between outbursts ranged from 26 to 124~days, and there appears to be a correlation between the time elapsed and ejecta mass (or rather cross sectional area).  We attempted to account for the correlation with the amount of insolation received at the surface by a single outburst source, but our simplified model could not adequately explain the correlation.  More information about the geological or topographic circumstances, and the mechanism(s) of the outbursts may be needed to further consider this correlation.

  The mini-outbursts of comet 67P are linked to steep scarps and cliffs, and in some circumstances can be directly connected to the collapse of such features (\citealt{vincent19-activity}, and references therein).  Extending this relationship to the mini-outbursts of comets 9P/Tempel~1, 103P/Hartley~2, and 46P, suggests that 46P has fewer cliffs per area than 67P and 9P, and is more similar to 103P.  This comparison is in agreement with the evolutionary sequence of \citet{vincent17-evolution}, which is based on a correlation between low topographical relief and insolation on the surface of 67P.

  Future studies of mini-outbursts and their relationship to topography would help us understand cometary behavior and nuclear surface evolution.  Comet 81P/Wild~2 potentially has frequent mini-outbursts, but observational circumstances from the Earth are less favorable for discovery than 46P at the time of our study.  However, comet 364P/PanSTARRS may present an opportunity to study mini-outbursts in 2023 ($\Delta\geq0.12$~au).  Furthermore, all spacecraft missions to cometary nuclei should consider observational campaigns dedicated to outburst discovery.

  \acknowledgments
  We thank D.~Schleicher and M.~Knight for contributing some Lowell 0.8-m telescope data, and B.~Skiff and L.~Wasserman for assisting with the Lowell observations.  We appreciate the helpful comments from the manuscript referees.

  Support for this work was provided by the NASA Solar System Observations program (80NSSC20K0673), the Space Telescope Science Institute (HST-GO-15372), the National Science Foundation (PHY-2010970), the National Research Foundation (NRF; No. 2019R1I1A1A01059609), the MINEDUC-UA project ESR1795, the European Union H2020-MSCA-ITN-2019 under Grant no. 860470 (CHAMELEON), and by the Novo Nordisk Foundation Interdisciplinary Synergy Program (NNF19OC0057374).

  This work is based on observations obtained with the Samuel Oschin Telescope 48-inch at the Palomar Observatory as part of the Zwicky Transient Facility project. ZTF is supported by the National Science Foundation under Grant No. AST-1440341 and a collaboration including Caltech, IPAC, the Weizmann Institute for Science, the Oskar Klein Center at Stockholm University, the University of Maryland, the University of Washington, Deutsches Elektronen-Synchrotron and Humboldt University, Los Alamos National Laboratories, the TANGO Consortium of Taiwan, the University of Wisconsin at Milwaukee, and Lawrence Berkeley National Laboratories. Operations are conducted by COO, IPAC, and UW.

  This work is also based on observations obtained by the MiNDSTEp team with the Danish 1.54m telescope at ESO’s La Silla Observatory.

  This research is also based on observations made with the NASA/ESA Hubble Space Telescope obtained from the Space Telescope Science Institute, which is operated by the Association of Universities for Research in Astronomy, Inc., under NASA contract NAS 5–26555.

  This research made use of Montage. It is funded by the National Science Foundation under Grant Number ACI-1440620, and was previously funded by the National Aeronautics and Space Administration's Earth Science Technology Office, Computation Technologies Project, under Cooperative Agreement Number NCC5-626 between NASA and the California Institute of Technology.

  This research has made use of data and services provided by the International Astronomical Union's Minor Planet Center.

  \facilities{PO:1.2m (ZTF), LO:0.8m, Danish 1.54m Telescope, HST (WFC3)}

  \software{astropy \citep{astropy18}, sbpy \citep{mommert14}, astroquery \citep{ginsburg19-astroquery}, JPL Horizons \citep{giorgini96}, Aperture Photometry Tool \citep{laher12-apt}, SEP \citep{barbary16-sep}, Ginga \citep{jeschke13-ginga}, DS9 \citep{joye03-ds9}, imexam \citep{sosey17-imexam}, astroscrappy \citep{curtis18-astroscrappy}, Montage \citep{jacob10-montage}}, photutils \citep{bradley21-photutils1.02}, calviacat \citep{kelley19-calviacat}

  \bibliography{journals,references,new-references}
  \bibliographystyle{aasjournal}
\end{CJK*}

\end{document}